# General Astrophysics and Comparative Planetology with the Terrestrial Planet Finder Missions

**Edited by:**
**Marc J. Kuchner**

***Based on a workshop held at Princeton University***
***14–15 April 2004***

**Dept. of Astrophysical Sciences**
**Peyton Hall**
**Princeton University**
**Princeton, NJ 08544**

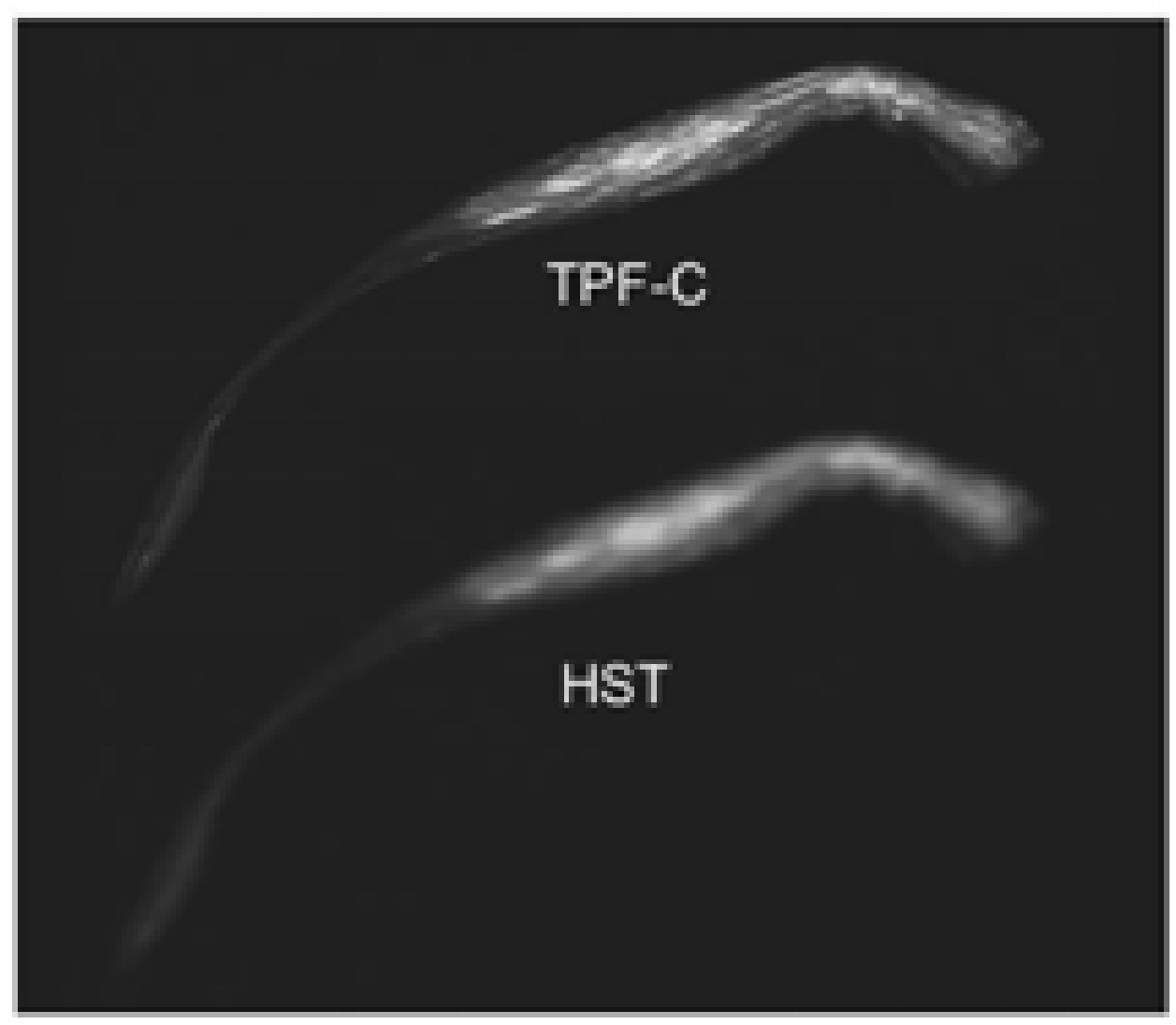

**Cover Illustration:** A galaxy lensed by a clumpy dark matter halo: simulated images from HST and from a wide-field camera on TPF-C. Because it can resolve HII regions in the lensed galaxy, the TPF-C image provides at least 4 times as many constraints on the lensing halo's mass distribution (Simon Dedeo and Ed Sirko).

# Contents

# 1 Introduction

The two Terrestrial Planet Finder (TPF) missions aim to perform spectroscopy on extrasolar Earths; TPF-C will operate in visible light, and TPF-I will operate in the mid-infrared. Extrasolar Earths are assumed to be roughly 26 magnitude in V band, roughly 0.3 μJy in the mid-IR, and located as close as roughly 30 milliarcseconds from a reasonable set of target stars, demanding high sensitivity, angular resolution and dynamic range to study. With capabilities matched to this task, the TPF missions could easily undertake a broad range of further scientific investigations.

This document discusses the potential of TPF for general astrophysics beyond its base mission, focusing on science obtainable with no or minimal modifications to the mission design, but also exploring possible modifications to TPF with high scientific merit and no impact on the basic search for extrasolar Earth analogs. It addresses both TPF-C and TPF-I, but emphasizes TPF-C, because its launch is planned for 2015, while TPF-I's nominal launch date is in 2019. This document does not attempt to describe all of the astrophysics TPF will be capable of, only to present some highlights of a meeting held at Princeton University, April 14–15, 2004, and some discussions that followed from that meeting. The participants in the meeting and in these discussions are listed in the Appendix.

In the next decade, the Atacama Large Millimeter Array (ALMA), the James Webb Space Telescope (JWST), and 20-m ground-based telescopes should begin operation, offering high sensitivity in the optical, IR, and submillimeter, and high resolution (~20 milliarcseconds) in the near IR and submillimeter. However, TPF-I and TPF-C will have special capabilities beyond the reach of these telescopes, as illustrated in Fig. 1. The TPF missions will offer high contrast imaging capability needed for the general study of extrasolar planetary systems. The stability of these space platforms could potentially be harnessed for astrometry of faint sources at the 100 milliarcsecond level. Furthermore, TPF-I will have unparalleled angular resolution in the wavelength range of 3.0–5.0 μm, and TPF-C will have unparalleled high-resolution imaging capabilities in the range of 0.3–1.0 μm. Our study emphasizes science that demands these special capabilities.

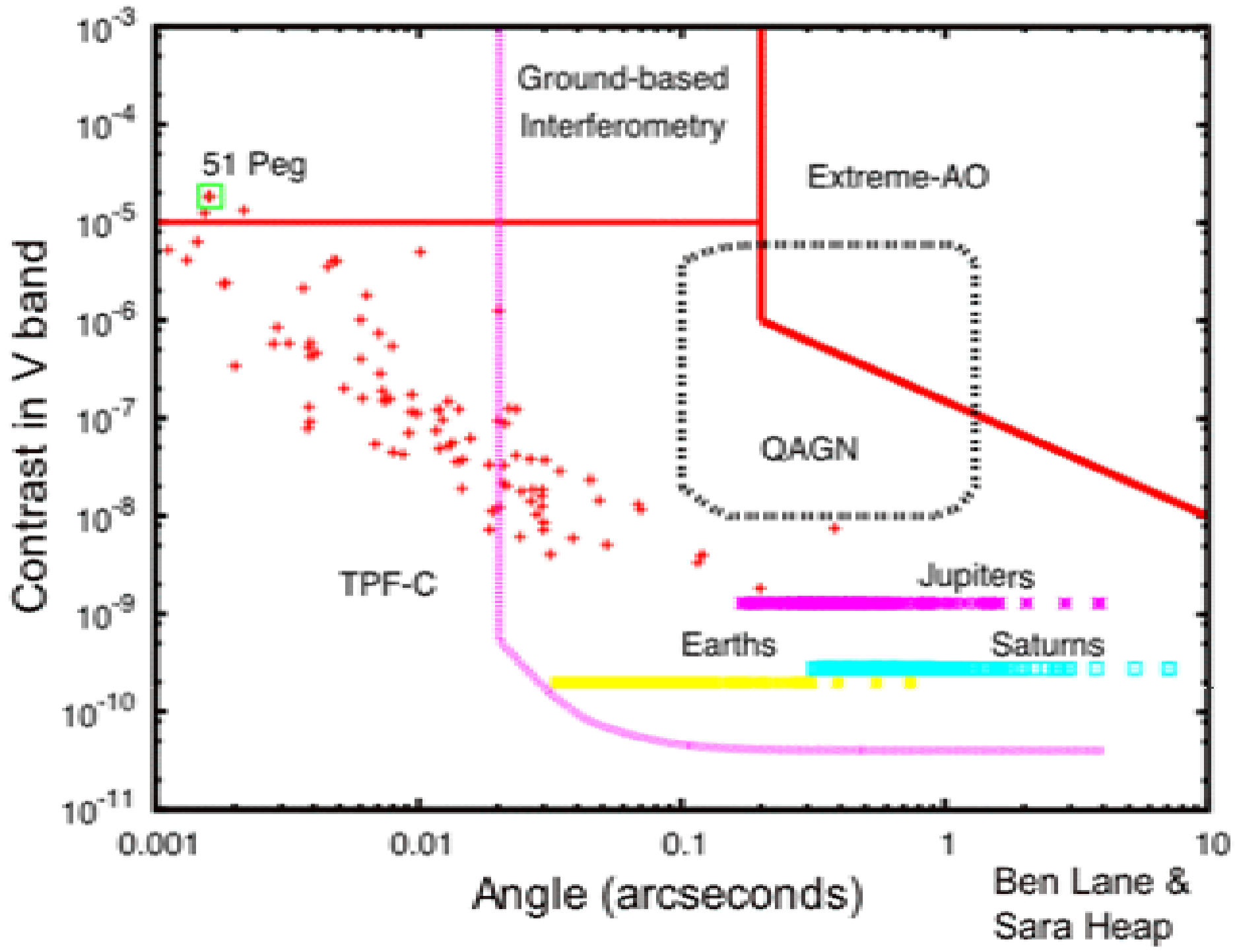

**Figure 1.** Parameter space uniquely accessible to TPF-C. The purple curve shows the contrast of TPF-C as a function of angular separation from a target star. Crosses indicate known precise-Doppler planets. The dotted black box indicates AGN host galaxies. The orange lines show the parameter space potentially accessible from the ground via interferometry and extreme AO on existing telescopes.

# 2 Comparative Planetology

*Sara Seager and Marc Kuchner*

The TPF missions are designed for the purpose of studying extrasolar planetary systems; their high-contrast, high angular resolution capabilities make them many orders of magnitude more powerful than any other planned missions for this purpose. We include a discussion of comparative planetology in this report because it has not been addressed elsewhere. By *planetology* we mean the study of all the components of planetary systems, including planets, small body belts, and dust; by *comparative planetology,* we mean considering a planetary system's components collectively and making comparisons among different planetary systems, including the solar system.

Comparative planetology might best be considered TPF's core mission! For example, on secular time scales, all the planets in a planetary system are coupled like masses on a string; the presence of a planet on an eccentric orbit anywhere in a system can cause an otherwise habitable-looking terrestrial planet to have eccentricity variations that take it far from the habitable zone every 10,000–100,000 years. Characterizing the terrestrial planets discovered by TPF will require taking an inventory of the other planets they orbit with.

Moreover, TPF can deliver a catalog of measurements on a reasonably large sample of planets ranging from terrestrial to giant, determining their radii, temperatures, and albedos. Combined with precise Doppler data and planet masses measured by SIM or TPF astrometry, this set of parameters indicates the densities and bulk compositions of these bodies, telling us whether we are looking at a gas giant, made mostly of H, a terrestrial planet made of Si and O, an ice giant, or perhaps some other beast with entirely different chemistry. These parameters can also indicate whether or not a giant planet possesses a core, suggesting whether it formed via gravitational instability or core accretion. If our theoretical understanding of planet cooling rates holds up, the measured temperature of a planet can constrain its migration history and indicate energy release from ongoing gravitational contraction and He rain. Low-resolution spectra will provide clues about the compositions and structures of planetary atmospheres. By measuring the temperatures and densities of extrasolar planets with TPF, we can take a giant step towards understanding the formation of planetary systems in general and the chemical and dynamical environment that produced life on Earth.

# 2.1 A Parallel Giant Planet Search

To understand an entire planetary system, giant planets must be detectable over a wide range of circumstellar separations. Jupiter and Saturn analogs in wide orbits around stars as old as the solar system are not detectable with any existing or planned ground-based or space-based telescopes other than TPF. Since a survey for Jupiter and Saturn analogs would require the same integration time as a survey for Earth-like planets and target roughly the same star list, and since we will need to find the giant planets in any systems containing terrestrial planets to assess their habitability, we recommend that the basic TPF design permit these surveys to be carried out in parallel with the survey for exo-Earths. A Jupiter-sized planet 10 AU from a star yields about the same flux as an Earth-sized planet 1 AU from a star, making this distance ratio—corresponding to an outer working angle of 10× the inner working angle—a natural outer working distance for a parallel giant planet survey. The brighter targets uncovered during the parallel planet search can be followed up with photometry and spectroscopy, as described in the next section.

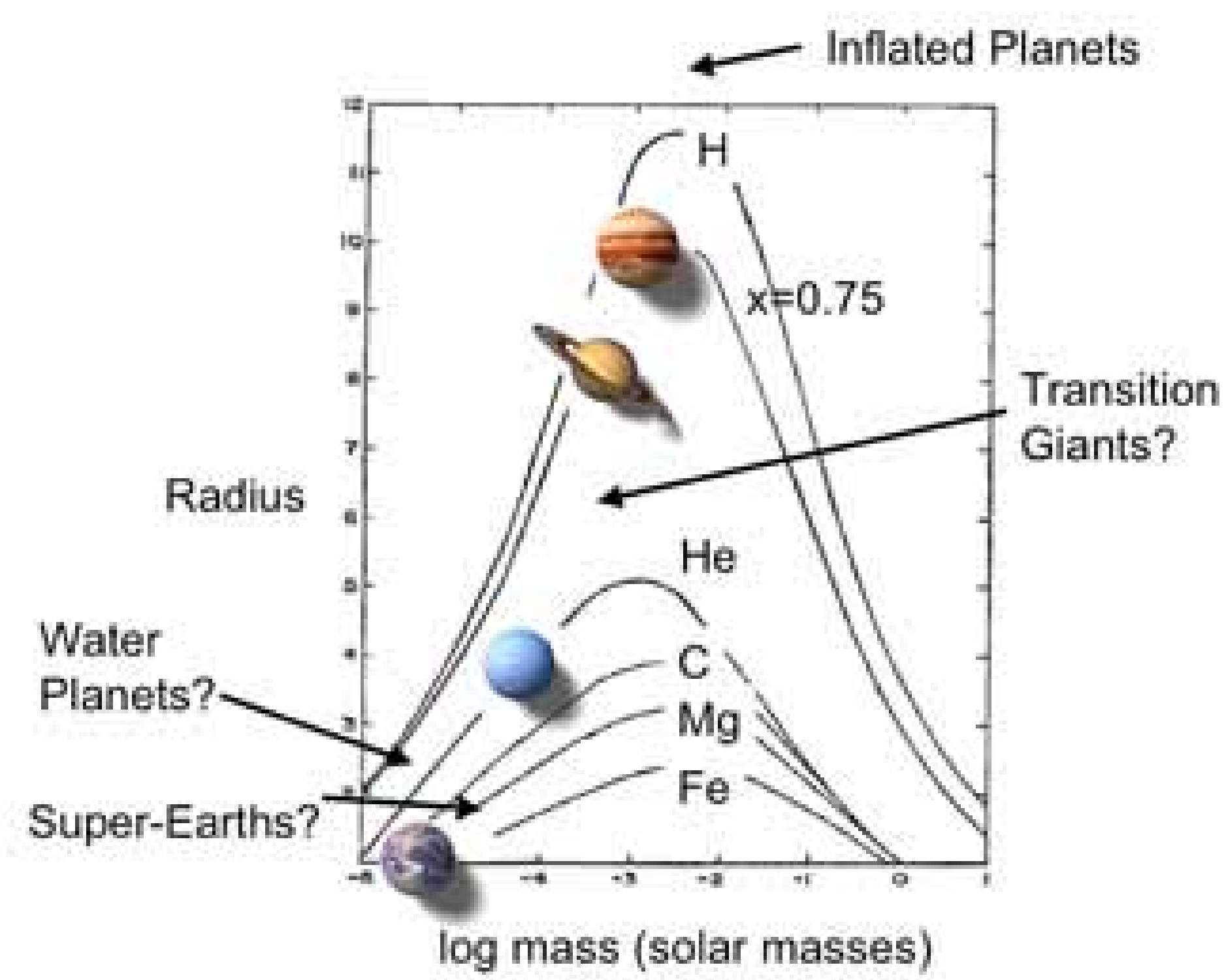

**Figure 2.** Fundamental diagram of comparative planetology, showing the masses and radii of Jupiter, Saturn, Uranus, and the Earth compared to the radii of zero-temperature spheres with various mean atomic weights (Zapolsky & Salpeter 1969). TPF combined with SIM will probably populate this diagram with a wide variety of extrasolar planets.

## 2.2 Already Known Planets: Radii, Temperatures, Masses and Albedos

Radial velocity surveys and the Space Interferometry Mission (SIM) will provide a wealth of giant and intermediate mass planets for study with TPF. Radial velocity surveys have already discovered well over 100 giant planets orbiting nearby stars, ~25 of which are accessible to TPF-C, as shown in Fig. 1. SIM will determine the actual masses and orbits of many of these planets. The SIM planet search list also contains the proposed TPF target star list. In TPF's core search region (> 0.8 AU), SIM will be able to detect planets down to ~0.1 $M_J$ out to 30 pc or ~1 $M_J$ out to 300 pc. TPF detection of planets measured by SIM can also help break degeneracies in the SIM astrometry and allow lower and/or more accurate masses to be determined. We suggest that a year of TPF time should be devoted to targeting known giant and intermediate-mass planets at favorable positions in their orbits.

The size of a planet is key to many of its properties, including its albedo and thermal emission. A measured radius is especially important for classifying new kinds of planets such as super Earths, ocean planets, and planets around non-solar type stars (e.g., Kuchner 2003). Figure 2 shows the masses and radii of Jupiter, Saturn, Uranus and the Earth, compared to the mass-radius relationships for zero-temperature spheres calculated by Zapolsky and Salpeter (1969). Because of these relationships, which probably apply to all but the hottest planets, a planet's bulk density indicates its mean molecular weight, which indicates its composition.

Making use of the mass-radius relationship in Figure 2 requires a robust measurement of a planet's radius, a measurement derived by detecting the same object at both visible and mid-IR light. Visible-light measurements of planet flux provide a combination of planetary albedo and area. Mid-infrared measurements of planet flux provide a combination of planetary temperature (from thermal emission) and area. Combining visible and mid-IR measurements breaks the degeneracy in a single-band measurement, giving the planet's size, albedo and radius; both TPF-C and TPF-I are required to understand the most basic characteristics of extrasolar planets.

The discoveries of Pluto and Sedna illustrate the importance of having both IR and visible light photometry to establish a planet's size. After its discovery in 1930, Pluto was estimated to be Earth-sized based only on visual observations and an assumed surface reflectance. This estimate was reduced when Pluto's icy nature was guessed. Finally the Charon-Pluto eclipses during the late 1980s constrained Pluto's radius to be much smaller—0.18 Earth radii. Sedna is a recently discovered small body in a highly elliptical orbit; it is currently 90 AU from the Sun. A non-detection of Sedna by *Spitzer* combined with visible-wavelength photometry constrains its radius to between roughly 1100 and 1800 km; a more accurate size determination will require an infrared detection.

Giant planet mass, radius, temperature and albedo are all important for characterizing extrasolar planets. As we mentioned, mass and radius are the key constraints on planetary bulk composition. For massive planets and young planets, mass and radius are also key constraints on planet evolution. Planets are born hot with a large radii, and contract and cool as they age (e.g., Baraffe et al. 2003; Burrows et al. 2003). Hence, a planet's radius can tell us about its evolutionary history, if the planet has a large unaccounted for interior energy source, and possibly whether or not the planet has a core (e.g., Bodenheimer et al. 2003). The well-studied transiting extrasolar giant planet HD209458b provides an example the importance of radius measurements; the measured radius at 1.42 Jupiter radii (Cody and Sasselov 2002) shows that the planet is composed predominantly of H and He. Yet theoretical evolutionary calculations (without an added interior energy source) disagree with the observed radius by 20–30%, a new puzzle in the field of planetary interiors. Evolutionary calculations also predict planet luminosities as a function of mass, radius and age. Compared with a measured temperature by TPF-I, these parameters will allow us to identify planets that, like Jupiter and Saturn, generate much of their own luminosity from sources such as gravitational contraction, tides, He rain, or other slow differentiation processes. The planetary albedo at visible wavelengths measured by TPF-C can indicate the heat balance of the planet and tell us the about the presence or absence of clouds.

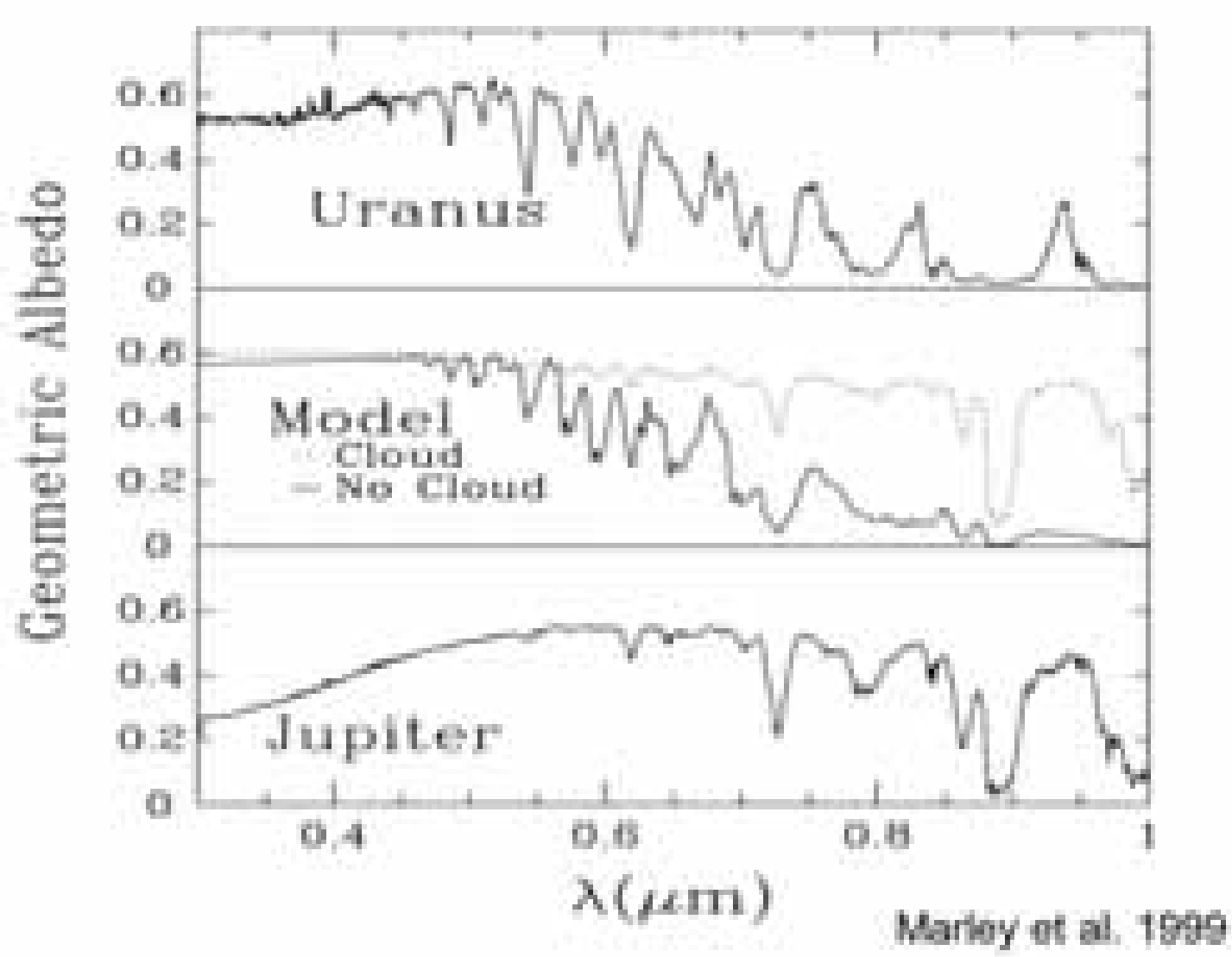

**Figure 3.** Synthetic giant planet spectra by Marley et al. (1999) showing how clouds affect the spectra of these planets. Without clouds, the model resembles the spectrum of Uranus (top); with clouds, the model resembles Jupiter's spectrum (bottom). Low-resolution TPF-C spectroscopy will probe the compositions of extrasolar planet atmospheres and identify cloudy planets, providing a first handle on planetary temperatures, albedos, and radii.

## 2.3 Low-Resolution (R = 100) Spectroscopy of Extrasolar Planets

TPF offers a unique opportunity to study a wide variety of planets including those with no solar system analogs. TPF, SIM and ongoing ground-based planet searches will provide a large sample of nearby extrasolar planets spanning a large range of sizes and masses. What are the atmospheric compositions of small giant planets and super Earths? Do the giant planets that orbit stars unlike the Sun have similar atmospheric characteristics compared to those in the solar system? Low-resolution spectra can begin to address these questions. Twenty-five of the known precise-Doppler planets are within reach of TPF-C spectroscopy, providing a ready-made target list for spectroscopy even before any new planets are discovered.

Low-resolution spectra can tell us a great deal about the structures and compositions of extrasolar planet atmospheres, even for planets without measured masses or radii. Model spectra for generic planets of a variety of masses and ages have been computed at a range of wavelengths (see Marley et al. 1999; Burrows et al. 2003; Burrows et al. 2004). However, much more work needs to be done to determine what physical parameters can be extracted from atmospheric spectral features accessible at TPF's low signal-to-noise ratio and low spectral resolution (R~70 at visible wavelengths; R~20 at mid-IR wavelengths). These physical parameters (including the presence and composition of clouds, the mixing ratios of the atomic and molecular species, and atmospheric structure) are sometimes degenerate; how well a TPF spectrum can constrain these physical characteristics still needs to be worked out to fully ascertain TPF's capability for giant planet investigation.

If extrasolar planets are like solar-system planets, low-resolution spectroscopy at visible wavelengths should be a good indicator of clouds. Figure 3 shows optical spectra of Uranus and Jupiter (taken from Karkoschka 1994) with their prominent methane absorption lines. It also shows synthetic spectra from a model by Marley et al. (1999), with and without optically thick ammonia clouds (middle panel). Clouds are responsible for much of the difference between the spectra of Uranus and Jupiter; with clouds, the model resembles the spectrum of Jupiter, and without clouds it resembles the spectrum of Uranus. Ammonia condenses much deeper in the atmosphere of Uranus (which also has an overlying optically-thin cloud deck of methane) than in the atmosphere of Jupiter; for a given planetary temperature, clouds indicate the temperature structure of a planet's atmosphere

Other parameters that can be derived from low-resolution spectra include planetary composition, (from e.g., signatures of $H_2O$, $CH_4$, alkali metals, and Rayleigh scattering at visible wavelengths, and signatures of $CH_4$, $NH_3$, $H_2O$ at mid-IR wavelengths) and estimates of surface gravity. Metallicity and photochemistry also play a role. Detecting rings and estimating their size with TPF-C may also be possible with spectra for planets with known orbits since rings are expected to have flat spectra whereas the planet has spectral absorption features.

If TPF-C or TPF-I observes a given planet, but not both, low-resolution spectroscopy can provide a first estimate of the planet's size, its temperature and thereby circumstellar distance by detecting some of the molecular species listed above. Different spectroscopic features are present at different temperatures; for example, the hottest planets will likely show alkali metals, cooler planers will conatin $H_2O$ vapor features, while on planets as cold as Jupiter, almost all water has condensed into clouds, leaving $CH_4$ and $NH_3$. Also, the strong $CH_4$ bands present on all solar system giant planets are not likely to be nearly as strong on terrestrial planets, where those species have been photodissociated, and the H escaped to space.

## 2.4 Characterizing Transiting Planets with TPF

TPF has the potential to make important follow-up measurements of known giant and intermediate-mass extrasolar transiting planets during both primary and secondary eclipse if high cadence measurement of bright stars (V = 9–16) is possible. Difference measurements before and after eclipse allow transit measurements using relative photometry, not absolute photometry. Hundreds of transiting planets are expected to be known before TPF launch from both ground-based and space-based programs. TPF-C offers a small advantage over JWST for these studies: access to the strong sodium line at 0.589 µm present in hot Jupiters. If TPF-C's bandpass extended blueward to 0.4 µm, it could also measure Rayleigh scattering, a useful indicator of the absence of clouds. TPF-I may also be useful for transit measurements.

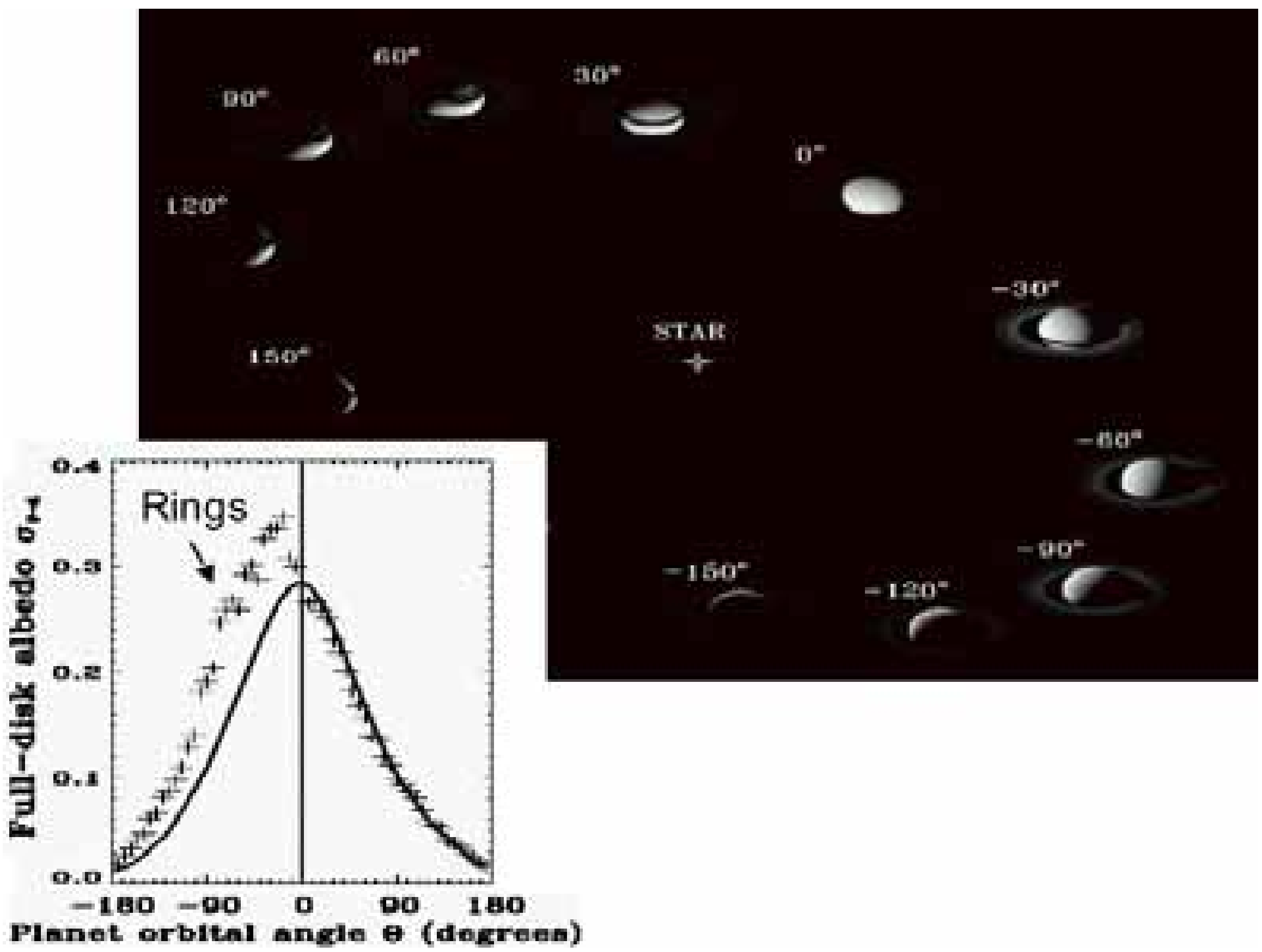

**Figure 4.** Phase offset in reflected light curve indicates rings (Dyudina et al. 2005).

## 2.5 Rings, Rotation, "Weather"

Synoptic visible and IR photometry of extrasolar planets can map their phase curves and their diurnal variability, indicating their seasonal weather and their rotation rates and potentially constraining the composition and size distribution of cloud particles. Planets with eccentric orbits may have exotic seasonal weather patterns. Visible light phase curves can also identify rings around planets with known orbits. Figure 4, from Dyudina et al. (2005), illustrates how Saturn's rings affect its phase curve; the phase curve including the rings (crosses) is brighter and offset in phase from the curve without the rings (solid line).

## 2.6 Asteroids, Comets and Volatiles

Although many missions claim to be able to study exozodiacal dust, TPF-C and TPF-I are the only planned missions that can detect and study true analogs of the solar-zodiacal cloud—a dust cloud located mostly interior to 4 AU with a face-on optical depth of $10^{-7}$. By simply determining the gross radial structure of exozodiacal clouds down to these faint limits, TPF can provide a deep census of extrasolar asteroid belts. Together with an inventory of massive planets and terrestrial planets and rudimentary spectra, the gross structure of exozodiacal clouds will help us understand the relationship between planets and asteroid belts and the delivery of volatiles to terrestrial planets.

The asteroid belt may have formed where it is because of the nearby presence of Jupiter stunted the growth of planet-sized bodies in this region. TPF can test this theory by studying the correlation between exozodiacal clouds and giant planets. Low resolution spectroscopy of dust clouds at ~10 μm can distinguish crystalline silicate dust like that seen in the coma of comet Halley from amorphous grains like those in the bulk zodiacal cloud.

The cosmic origin of volatile condensates on the Earth is an unresolved question, and an important one, since these condensates include the water, carbon, and nitrogen in our own bodies. Did the Earth's water precipitate directly from the solar nebula, or was it delivered later by giant primordial asteroids? Were comets or asteroids more important volatile delivery agents? Modeling the spectra and images of extrasolar planetary systems mapped by TPF can answer these questions.

# 2.7 Useful Extensions to TPF for Giant Planet Science

## 2.7.1 TPF-C

Extending the bandpass to 1.5 μm would allow TPF-C to reach a strong water band that could help TPF find fainter giant planets. Extending the bandpass to 1.7 μm would allow it to reach a strong methane band that would indicate the overall oxidation state of a planet's atmosphere and possibly reveal methanogenic life. Extending the band to 0.4 μm or shorter would probe atmospheric Rayleigh scattering and indicate total atmospheric column depth. This extension would also be useful for studying atmospheric photochemistry, a process that strongly affects the albedos and colors of Jupiter and Saturn in the visible.

## 2.7.2 TPF-I

Extending the bandpass from 6.5 to 17 μm would allow TPF-I to reach $H_2$ CIA features, helping it distinguish planets with massive amounts of $H_2$. Extending the band to 4 or 5 μm would allow TPF-I to reach the 5 μm peak in giant planets (the peak is at 4 μm for hot planets), as well as a very strong $H_2O$ absorption band.

# 3 Circumstellar Disks

*Marc Kuchner*

The solar system contains a variety of “debris” left over from planet formation—asteroids and comets and Kuiper belt objects—largely distributed in belts near the ecliptic plane. These small bodies continually produce dust as they collide and outgas. One out of every ~10 main sequence stars shows a strong far-infrared excess that we attribute to an dust in an extrasolar debris disk, produced by extrasolar comets and asteroids. These disks are the signposts of extrasolar planetary systems. Most known extrasolar debris disks are thousands of times brighter than the solar system; only TPF can detect a dust cloud faint enough to be real solar-system analog.

Disks around stars younger than 10 million years old, such as T Tauri stars and their more massive analogs, Herbig Ae stars, are often called protoplanetary disks because these disks are the likely sites of jovian planet formation. Protoplanetary disks are bright and optically thick, but the young stars that host these disks are typically several times farther away than planet search targets and debris disk targets. So studying these disks requires only moderate contrast ($\sim 10^5$), but benefits especially from high angular resolution that can resolve scales of <1 AU at the distance of the star-forming regions in Taurus (<7 milliarcseconds).

Dividing disks into debris disks and protoplanetary disks has already become too simple a scheme; by the time TPF flies, these classifications may be meaningless. Ordinarily, the line might be drawn where the amount of gas in the disk becomes undetectable. As protoplanetary disks lose gas, we can presently only follow them so far; for most young stars we are presently incapable of detecting disks in nearby star-forming regions less massive than a few Jupiter masses of gas, since less massive disks often produce only tiny excess emission. But quantities of gas down to 1 Earth mass may be crucial to terrestrial planet formation, causing terrestrial planets to migrate, damping their eccentricities, altering their formation mechanisms, and supplying them with proto-atmospheres and essential volatiles. TPF-C could help fill in the missing link in disk evolution by mapping optically-thin disks in metal emission lines, as Brandeker et al. 2004 and Liseau 2004 have done for β Pictoris, presently the only good example of a debris disk with gas. Such maps would show the distribution of gas in disks during the late stages of planet formation.

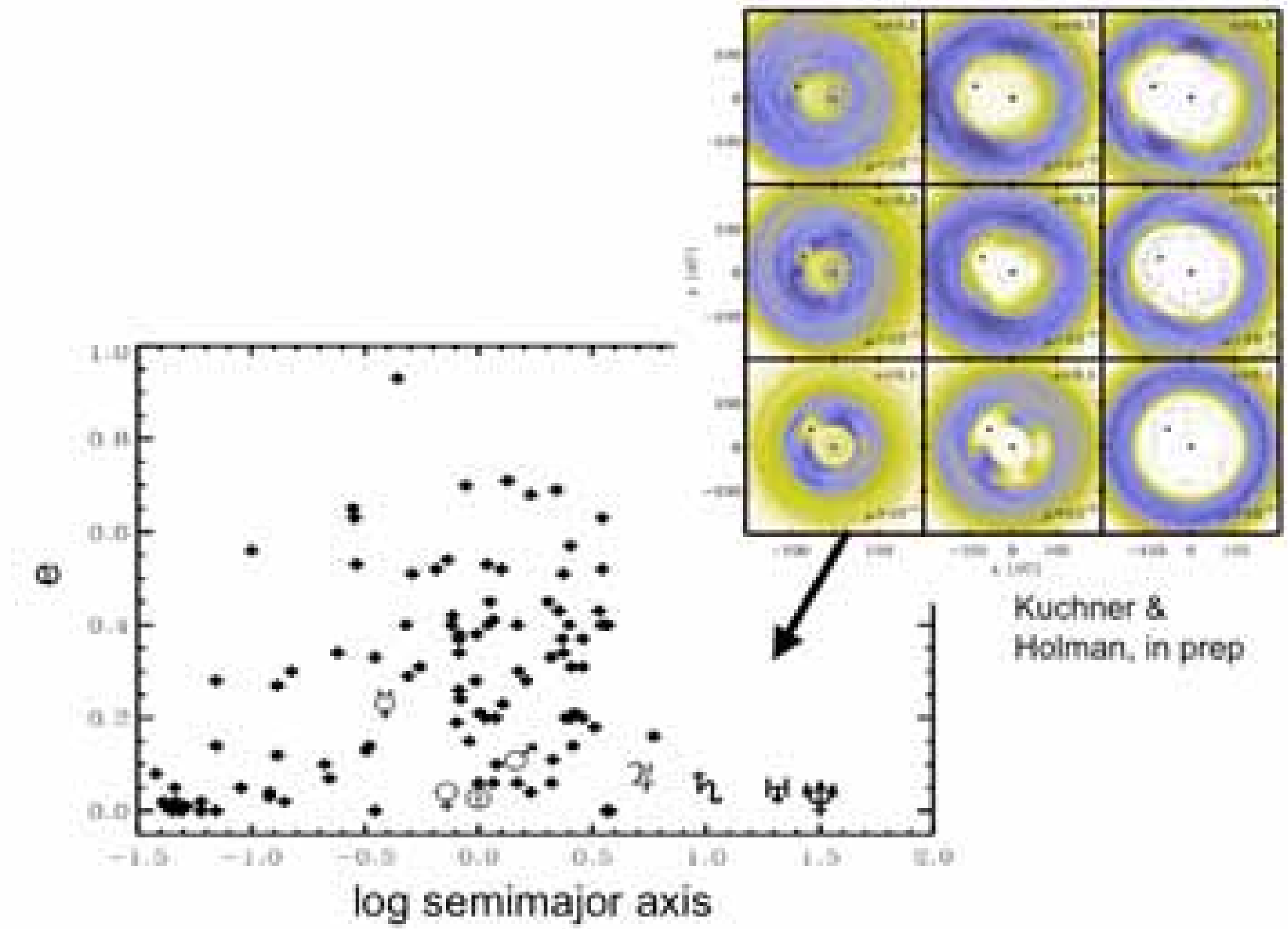

**Figure 5.** Orbital eccentricities and semimajor axes of extrasolar planets detected by the precise-Doppler method (see http://exoplanets.org) and models of collisionless dust disks containing planets of various masses and eccentricities (inset). Observing these structures with TPF can fill in the gap at > 10 AU, where the planets are too faint to detect directly, and orbital periods are too long for indirect methods, like the Precise Doppler method.

Studying the mass-distribution and chemistry of protoplanetary disks and debris disks at AU size scales is a natural task for ALMA. But TPF will be sensitive to dust grains too small to emit in the submillimeter. These small grains probably first intercept stellar radiation in optically thick disks, governing the energy balance of passive disks (e.g. Chiang & Goldreich 1997); TPF-C and TPF-I can provide unique observations of the structure of the disk photospheres. Moreover, debris disks TPF can image in an hour during the course of a planet search will require tens or hundreds of hours for ALMA.

A few more opportunities for disk study provided by TPF are described below.

# 3.1 Indirect Planet Detection

Planets orbiting at semimajor axes beyond roughly 10 AU have orbits too long to permit detection via radial-velocity or astrometric techniques in a human lifetime, and they will be too faint for TPF to detect in reflected light. But these coldest planets may tell critical chapters of the planet formation story. New models of planet-disk interactions (Type III migration) suggest that planets can easily migrate out to the outermost parts of protoplanetary disks (Masset & Papaloizou 2003). Is this mechanism the dominant migration mechanism? Or do planets mostly migrate inwards, as the discovery of 51 Pegasi-type planets suggested? The outermost planets can dynamically excite inner planets by dynamically coupling them to passing stars (Zakamska & Tremaine 2004). Could this mechanism explain the large eccentricities of the observed precise-Doppler planets? The recent announcement of a massive planet candidate found ~50 AU from a young brown dwarf in the TW Hydra association suggests that these far-out planets may be common (Schneider et al. 2005).

The only way to detect these most distant planets around stars of solar age and mass may be to study the structures of debris disks. Resolved images of extrasolar debris disks reveal resonant structures that probably indicate the presence of planets buried in the dust. Recent advances in debris disk dynamics (e.g., Kuchner & Holman 2003) allow us to use the structures we detect to infer the mass and eccentricity of a perturbing planet as small at ~10 Earth masses.

The present handful of submillimeter and far IR debris disk images only hint at the promise of this technique. The disks imaged so far are complex systems, 100–1000 times as optically thick as disks TPF can image, disks where collisions and gas complicate the dynamics. TPF will have the unique ability to image dust clouds as optically thin as the solar zodiacal cloud—relatively easy to model and interpret.

# 3.2 The Central 0.2 AU and the Origin of Close-In Giant Planets

Since TPF-I can potentially offer resolution of ~1 milliarcseconds in the near infrared, it can probe the innermost regions of protoplanetary disks, regions that defy standard disk models, according to ground-based optical interferometers (e.g., Millan-Gabet et al. 1999). The central regions of protoplanetary disks are credited with hosting a variety of mysterious phenomena: chondrule formation, launching winds, tall inner walls. These regions are also vital to understanding the origin of the many extrasolar planets we have observed in 1–6 day orbital periods; either planets naturally form in these hottest regions of the disk, or planet migration halts there (e.g. Kuchner & Lecar 2002).

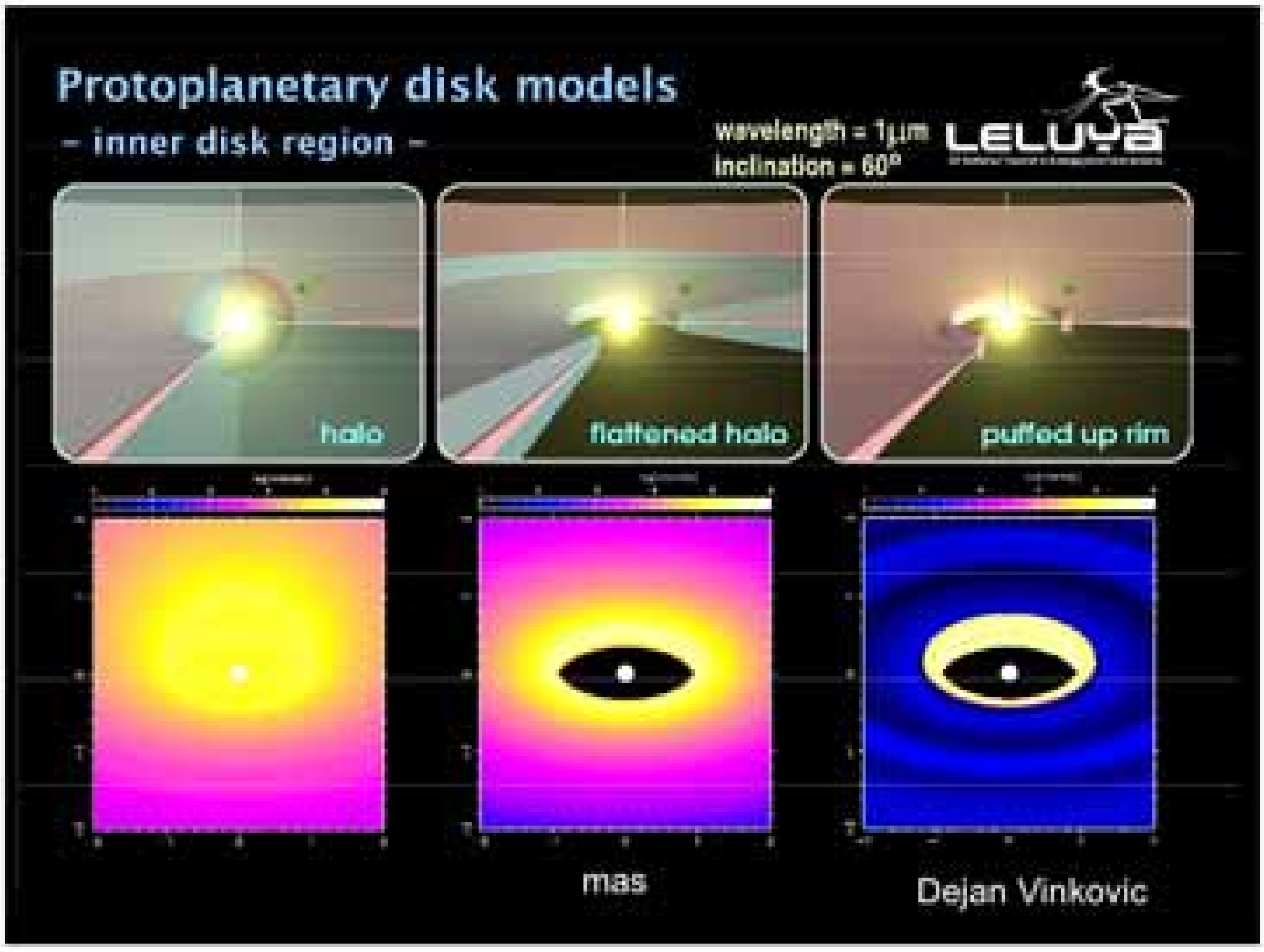

**Figure 6.** Three possible geometries for the centers of T Tauri disks and synthetic images of these models in scattered light. TPF-I could distinguish among these possibilities, shedding light on the origin of close-in planets (Dejan Vinkovic).

Figure 6 shows three possible geometries for T Tauri-disk centers: a spherical halo of small dust grains, a flattened halo, and a disk with a puffed up inner rim (e.g. Dullemond et al. 2001). The lower panels show radiative transfer models of disks by Dejan Vinkovic, with these geometries, showing how they would appear at 1 μm, shining mostly in scattered light. TPF-I could distinguish among these geometries, which have implications for the temperatures of the central disk regions, which determine the critical core masses of planets that can form there.

## 3.3 Molecular Hydrogen: The Dark Matter of Planet Formation

Most of the mass of protoplanetary disks is probably in the form of molecular hydrogen; since this substance is so difficult to observe, even ALMA will not map the true mass distribution of protoplanetary disks, only the distribution of rarer gasses like CO that are assumed to trace the distribution of the hydrogen. This approach is inadequate for studying photodissociation and photoevaporation of disks and disk chemistry. Although older stars are less likely to have massive disks, what mechanism depletes the disks is unknown. Although we can observe the chemistry of today's solar system, the current chemistry does not clearly indicate whether of not the disk ever reached chemical equilibrium, and hence we do not know the original distribution of volatiles, like water and carbon in the disk. Unraveling these mysteries probably requires observing the distribution of molecular hydrogen in T Tauri and Herbig Ae disks.

With a very high resolution spectrometer (R~100,000) TPF-I could map protoplanetary disks and debris disks in molecular hydrogen, the material that forms the bulk of the solar nebula and the bulk of gas giant planets. TPF-C could measure pure-rotational $H_2$ transitions, S(0) 28.0 μm, S(1) 17.0 μm, S(2) 12.3 μm, S(3) 9.7 μm and S(4) 8.0 μm using a high-resolution spectrograph (R~ 100,000). Since molecular hydrogen is expected to be the dominant species in these disks, measuring the quantity of molecular hydrogen in these disks will indicate the dynamical environment of terrestrial planet formation. Measuring both the 17- and 28- μm lines would allow determination of the $H_2$ mass. The J values of the transitions are in parentheses above; measuring two lines from both ortho- and para- states (even and odd J-values) would be a sensitive probe of the gas mass and formation temperature.

# 4 Stars, One at a Time

## 4.1 Stellar Orbits around the Galactic Center Black Hole

*Nevin Weinberg, Milos Milosavljevic, and Andrea Ghez*

Near-infrared monitoring with speckle and adaptive optics techniques with 10-m class telescopes has enabled complete orbital reconstruction of several stellar sources orbiting the ~4 $\times 10^6$ $M_{sun}$ black hole at the Galactic center (Schoedel et al. 2003; Ghez et al. 2003a). Some sources have been monitored with astrometric errors of a few milliarcseconds, and radial velocity errors < 50 km $s^{-1}$ (Eisenhauer et al. 2003; Ghez et al. 2003a), allowing the detection of the accelerated proper motions of ~10 stars, as illustrated in Fig. 7. The star with the shortest orbital period (S0-2) has an orbital period of ~15 yr.

The greatest obstacle to detecting and following new stars is stellar crowding in the central arcsecond. Light contamination from nearby bright stars and underlying faint stars floods the pixel elements, imposing a limit on the faintest detectable star. The higher the resolution of a telescope, the more Galactic center stars it can detect and monitor. With a wide field camera, TPF-C can provide 100 microarcsecond astrometry down to 23 magnitude in the near-infrared. This factor of ten improvement over current 10-m class telescopes will enable TPF-C to follow the orbital motions of more than 100 stars, with orbital periods in the range of 3–100 yr (Weinberg, Milosavljevic, & Ghez 2004, hereafter WMG; see Fig. 7). The larger number of monitored stars, the improved accuracy of their astrometric positions, and the detection of shorter period orbits, will provide a very precise probe of the Galactic center potential.

If the extended matter distribution enclosed by the orbits has a mass greater than approximately 1000 $M_{sun}$ at 0.01 pc from the black hole, it will produce deviations from Keplerian motion detectable with TPF-C (WMG). This dark-matter concentration is within the range of theoretical predictions (Gondolo & Silk 1999). The effects of dark-matter have never been detected on such small scales.

The IFU on TPF-C will provide radial velocity measurements of the stars in the field. The detection of 3D orbital motions would break the degeneracy between measurements of the black hole's mass and measurements of its distance, providing a measurement of the distance to the Galactic center $R_0$. TPF-C should be able to measure $R_0$ to better than 0.1%; when combined with solar proper-motion measurements from future astrometric surveys, this measurement will provide precise constraints on the Milky Way's dark-matter halo shape (WMG). The halo shape is a sensitive probe of dark-matter models and structure formation scenarios; currently it is very poorly constrained (Olling & Merrifield 2000).

The short-period orbits detectable with TPF-C will likely have pericenter passages that bring them within a few tens of an AU (~500 Schwarzschild radii) of the massive black hole. The lowest-order relativistic effects on the orbits, like prograde precession, will be detectable with TPF-C. Higher order effects like frame dragging to the spin of the black hole may also be within reach of TPF-C (WMG).

The stellar mass function inside the dynamical sphere of influence of the black hole is likely dominated by massive remnants, including stellar-mass black holes, ~20,000 of which are thought to lie within 1 pc of the central black hole (Miralda-Escude & Gould 2000). Perturbations from remnants should deflect stellar orbits, changing their orbital energies at a rate proportional to the mass of the remnants; monitoring of stellar proper motions can therefore be used to directly probe the mass function of stellar remnants. Over a ten-year baseline, approximately 10% of all galactic center stars monitored at TPF-C precision should undergo detectable encounters with background remnants if the remnants are stellar-mass black holes (Milosavljevic and Weinberg 2004, in preparation).

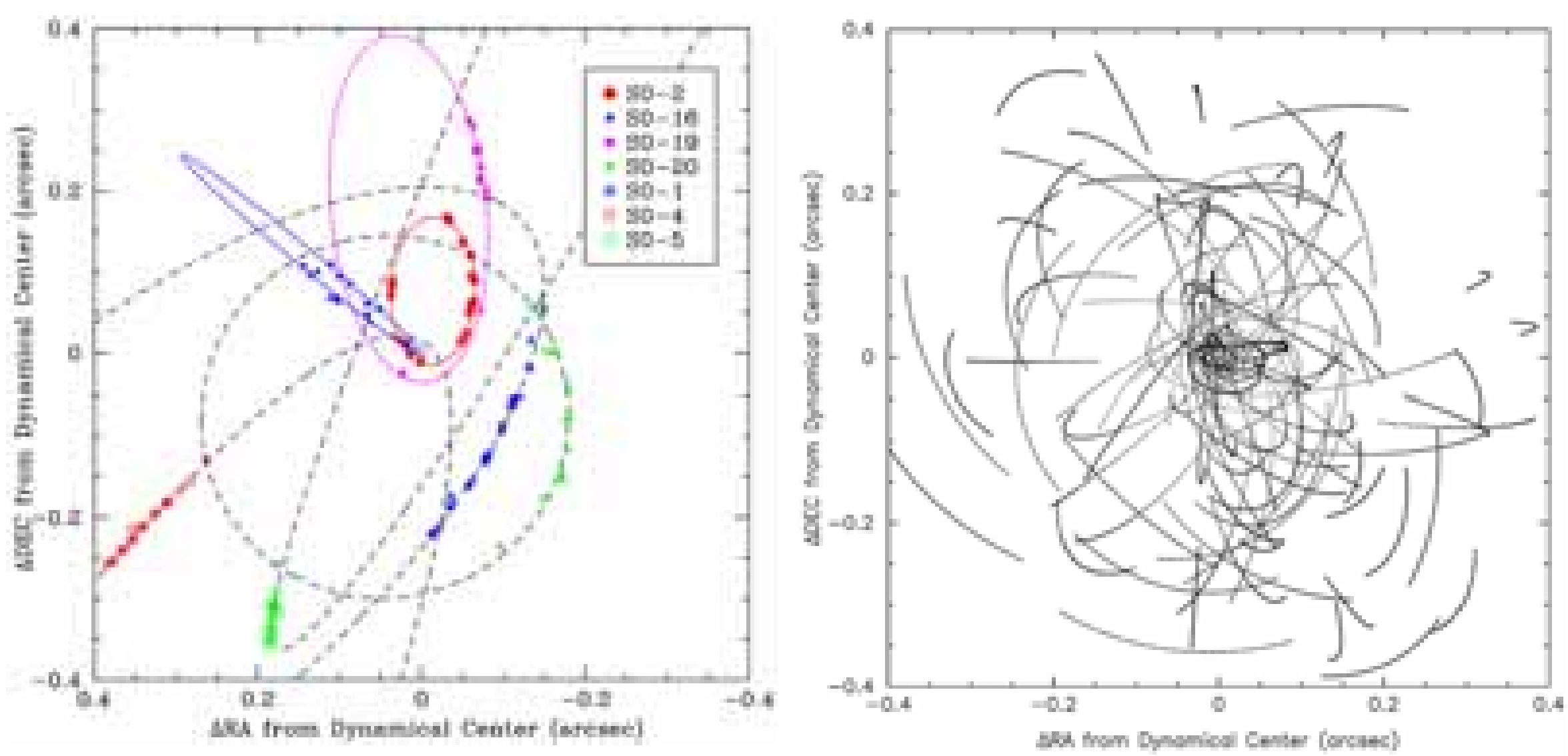

**Figure 7.** (Left panel) Astrometric positions and orbital fits for 8 stars that show accelerated proper motion within the central 0.8"×0.8" of the Galaxy (Ghez et al. 2003b). (Right panel) Astrometric positions for a simulated sample of 100 stars detectable with TPF-C. Motions are over a 10 yr baseline assuming ten observations per year per star.

## 4.2 Stellar Populations and Galactic Star Formation

*Thomas M. Brown*

Main sequence photometry provides the most direct method for determining the ages of star clusters and other stellar populations, telling us the star formation histories of these populations. The Hubble Space Telescope can slowly obtain optical photometry of the oldest main sequence stars out to the edge of the Local Group of galaxies (e.g., Brown et al. 2003), creating a color-magnitude diagram that would enable a reconstruction of the star formation history back to the Big Bang. The largest galaxy of the Local Group, Andromeda, shows surprising evidence of a violent merging history, including complex substructure and interaction remnants (Ferguson et al. 2002) and an extended star formation history in the halo (Brown et al. 2003). HST continues to probe the star formation history in these substructures, but can only examine a few more sightlines within its remaining mission.

However, the Local Group does not contain a representative sample of the universe; it contains no giant elliptical galaxies and only two giant spiral galaxies (the Milky Way and Andromeda). Beyond the Local Group, HST can only resolve younger (brighter) main sequence populations, truncating the history that can be measured. The Andromeda galaxy and most dwarf galaxies in the Local Group will remain largely unexplored in the remaining HST lifetime.

With a diffraction-limited optical telescope, the distance at which such stars can be measured increases linearly with aperture diameter and the exposure time needed to measure stars at a fixed distance decreases with the fourth power of aperture diameter. With an effective diameter ~3 times that of HST, TPF-C could measure star formation histories well beyond the Local Group and enable wider surveys of the Local Group populations using a wide-field camera. With 200 hours of exposure, the TPF-C could measure the star formation histories in galaxies nearly 4 Mpc away. The accessible volume of space encompasses well over 100 galaxies, including a giant elliptical and several more giant spirals in the Cen-A Group and M81-M82 Groups. Because the exposure time required for photometry within the Local Group drops by more than an order of magnitude, the TPF-C could also provide a more complete picture of the star formation histories in the Local Group, sampling not only the substructure of Andromeda, but also dozens of dwarf galaxies.

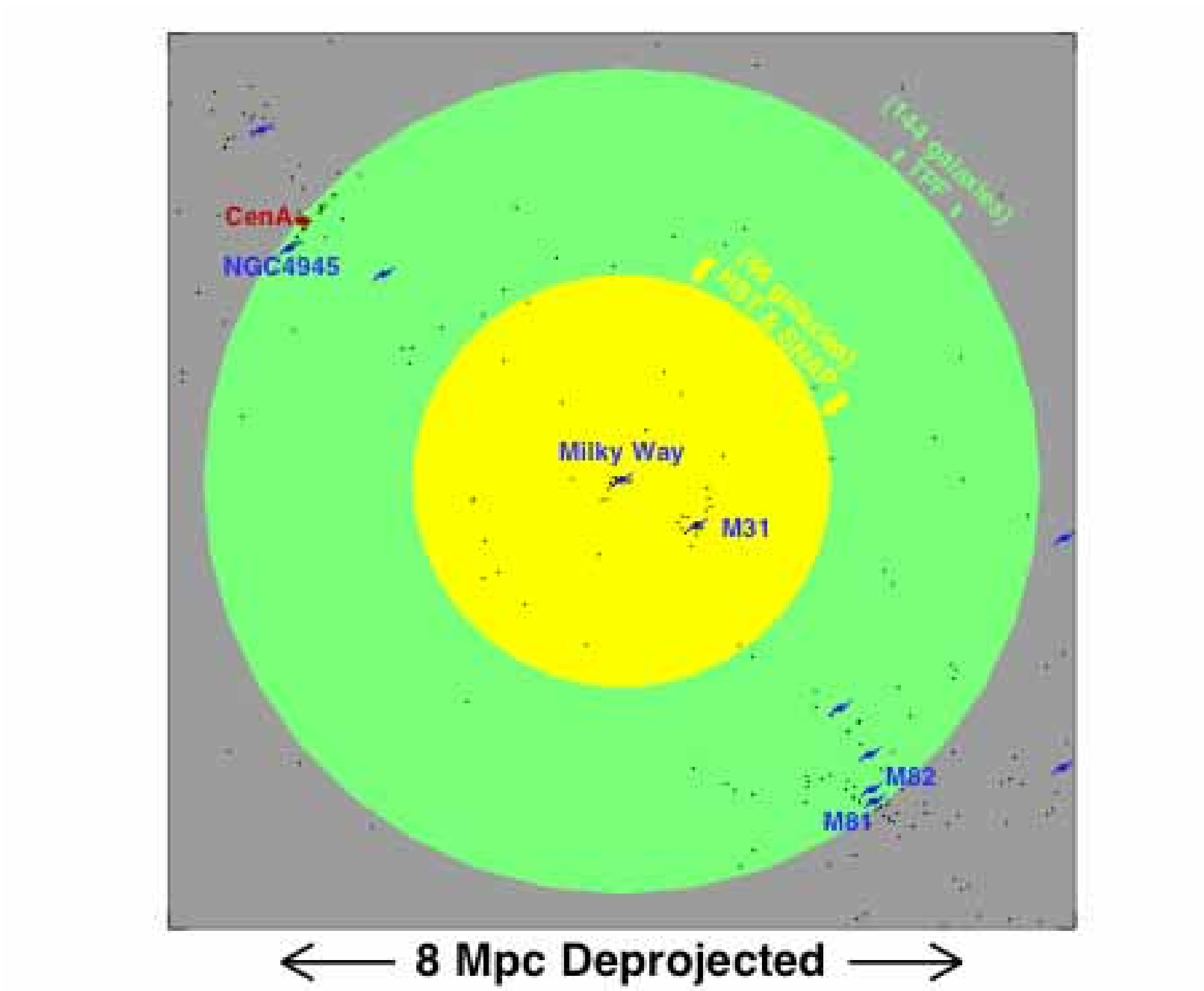

**Figure 8.** Deprojected view of the galaxies in the local universe based on Karachentsev et al. (2004) showing the volume that HST, SNAP, and TPF-C can survey photometrically down to the 12-Gyr main-sequence turn off in 200-hour broad-band integrations. Spiral and elliptical symbols indicate the giant galaxies. TPF-C can study stellar populations at nearly 4 Mpc.

## 4.3 AGB Stars and Planetary Nebulae

*Orsola De Marco, Bertrand Mennesson, & Dejan Vinkovic*

AGB stars and planetary nebulae, their progeny, drive the chemical evolution of the universe. Mira stars have the highest mass loss rate of the AGB, up to $10^{-4}$ solar masses per year, providing the largest fraction of the heavy elements eventually found in the interstellar medium. AGB winds supply the universe with its carbon, including the C-bearing molecules and grains in the ISM and pre-solar dust, the source of carbon for life on Earth.

These important sources are poorly understood. Although Mira variables have been well observed in wavelength regimes from the UV to the radio domain, basic physical parameters like the photospheric size, effective temperature, and mode of pulsation of these stars are uncertain to factors of ~2, as well as the actual mass loss and pulsation mechanisms (Willson 2000, Whitelock et al. 2000, Wood 2000). The physical processes that create and shape AGB outflows remain unknown.

### 4.3.1 Mira Variables

Large and bright, Mira stars make easy targets for TPF-I. With high-angular resolution observations over a wide range of wavelengths, TPF-I can simultaneously probe several regions of a Mira star's atmosphere. The 2–25 µm region is particularly informative as it covers the three main physical regimes of the stellar environment.

The shorter wavelengths make it possible to probe the actual photosphere, and TPF-I will provide 10x10 pixels maps for the largest Miras (R Leo, o Ceti, R Cas …). The 3–10 µm region is very rich in spectral features from gas molecules present in the upper atmosphere. This was already evidenced by measurements with the ISO SWS instrument, but at much lower spatial resolution than TPF-I, which still yields quite model dependent information (Yamamura et al., 1999). The longer wavelength range will allow accurate determination of the dust content and spatial distribution, including asymmetries. TPF-I can easily achieve a spectral resolution of 1000 or higher on such bright objects, enough to accurately characterize the chemical composition of the photosphere and close-in circumstellar regions.

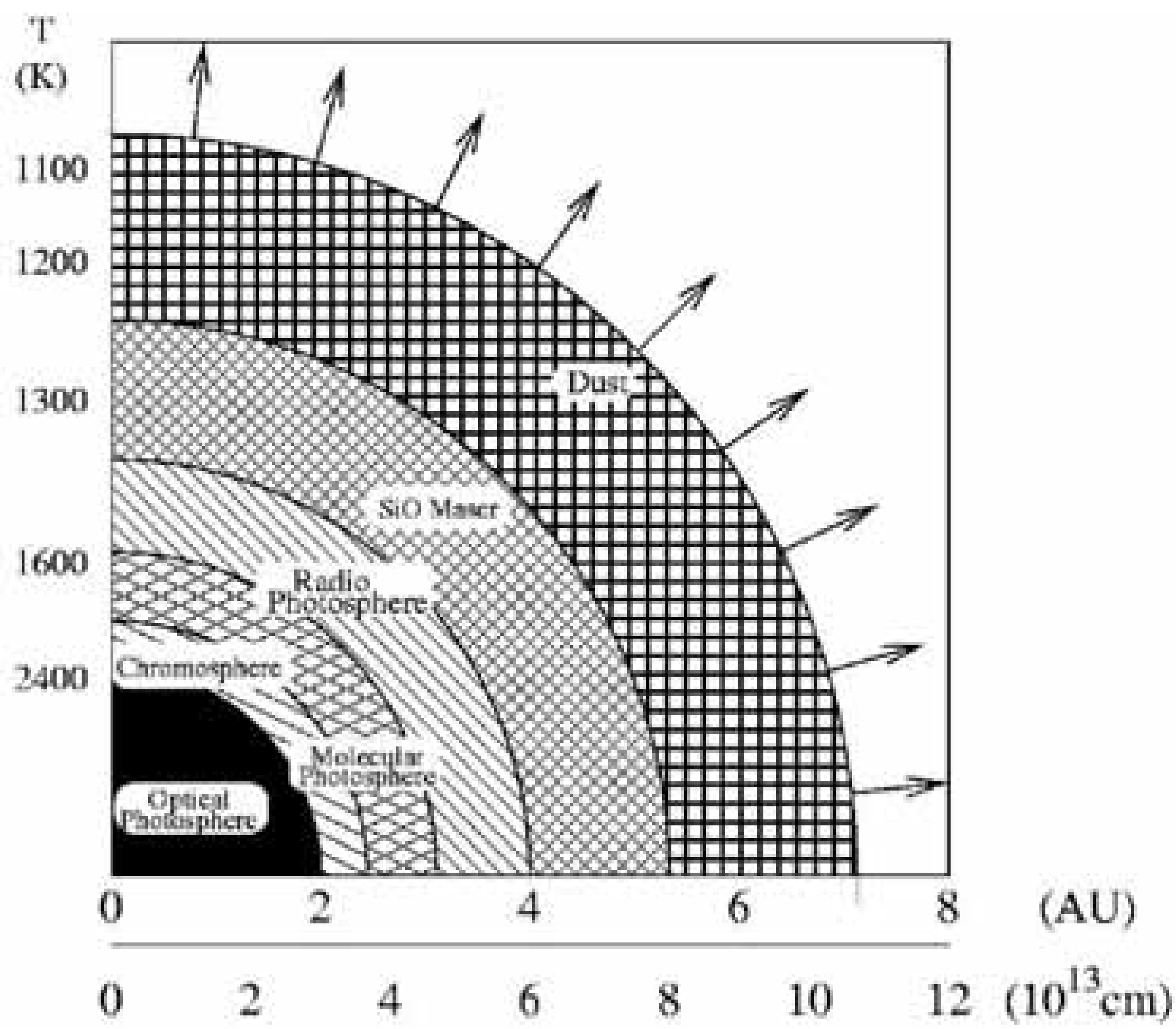

**Figure 9.** Schematic of the atmospheric spatial structure and temperature profile of a typical Mira star (Reid & Menten 1997).

### 4.3.2 Planetary Nebulae and the Origin of Carbon

One of the most intriguing problems of stellar astrophysics is the transition from spherically symmetric asymptotic giant branch (AGB) stellar winds to mostly asymmetric planetary nebulae. The initial phase of this transition appears in the end of AGB evolution, when a still unknown physical process initiates a high-velocity, low-density bipolar outflow. When the asymmetry expands to larger scales during the proto-planetary nebulae phase, it appears highly complicated in morphology, sometimes displaying multiple jet-like structures. It is suspected that companion stars or planets may play a crucial role in this phenomenon (De Marco et al. 2004). High-contrast TPF observations of planetary nebulae can search for hidden companions to the central stars of planetary nebulae, testing the role of binary companions in creating the nebulae and also the existence and survival of planetary systems around AGB stars, an intriguing problem related to the future of our own Solar System.

Mapping stars in the late-AGB to early post-AGB period in the 10 μm PAH and 20 μm oxygen-dust feature groups can allow us to quantify the mass that resides in circumstellar disks (O rich) and outflows (C-rich). These measurements can indicate whether the presence of a companion is enhancing the mass-loss, whether the amount of carbon dust produced is consistent with the current levels of carbon enrichment of the ISM. Correlations between position along the AGB, binarity, and dust type and distribution around the star will pin down how and when the binary intervenes in the evolution of the star.

There is a large discrepancy (up to a factor of ~20) between the PN abundances determined from collisional and recombination lines. These abundances are fundamental in determining the chemical evolution of galaxies as well as stellar yields for our galaxy. A way to reconcile this is to include in the models small H-deficient globules. These globules must be in most cases well below the detectability of HST. They are thought to be small, but whether they are promoted by the star or they are due to some chemical event in the circumstellar gas, it is unknown. Detecting these globules in planetary nebulae and measuring their compositions with TPF would be a fundamental step in understanding stellar yields from AGB stars.

TPF-C studies of planetary nebula formation would benefit from a large outer working angle (a few arcsec), a broad coronagraphic null for removing light from these large stars (~30 mas) and a near-IR camera. TPF-I would be a powerful tool for hunting for binary companions.

# 5 Galaxies and Active Galactic Nuclei

## 5.1 Galaxies

### 5.1.1 TPF-C Deep Fields

*David Spergel*

TPF-C will devote roughly 20 days of integration time per target to search 50 nearby stars for planets. If TPF-C is equipped with a wide-field optical/near-IR camera, then these long integrations will provide astronomers with ~25 low galactic extreme deep fields. These extreme deep fields, extending 4 magnitudes fainter (for sky background limited objects) than the Hubble deep field and covering 25 times the area at more than double the spatial resolution will be gold mines for astronomers studying galaxy formation and evolution.

The Hubble Deep Field and the Hubble Ultra Deep Field are not only striking astronomical images that have captivated the public, but are also powerful probes of galaxy evolution. The combination of HST imaging, spectroscopy from 6–10 meter class telescopes and far infrared observations of high redshift dust have revolutionized our view of the early universe. In the coming decade, the combination of TPF-C imaging at 0.4–1.7 μm, JWST diffraction limited imaging from 2–15 μm, spectra from the next generation of large telescopes (20–30 m) and ALMA's high resolution images of gas and dust in high redshift galaxies will enable cosmologists to understand the evolution of galaxies.

The combination of TPF-C high-resolution optical observations (rest-band UV) and JWST's near and mid-infrared observations will measure the stellar populations in high redshift galaxies. As in the comparative planetology studies discussed earlier, wide spectral coverage has a multiplicative effects in its ability to constrain galaxy model parameters. GALEX's recent observations show the power of ultraviolet imaging at tracing star formation in nearby stars. TPF-C will be able to carry out this program at higher redshift.

## 5.1.2 Young Emission Line Galaxies

*Bruce Woodgate*

A large fraction of galaxies in the early universe have emission lines, diagnostics of strong star formation and AGN activity. Observing the structure of these galaxies in redshifted O VI, Lyman $\alpha$, and C IV, as they form from the most concentrated dark matter cores, is the best way to estimate the bias factor between the observed distribution of galaxies and theoretical distribution of the dark matter. Assuming JWST will operate down to 0.6 μm and be diffraction limited at 2 μm, TPF-C will be the telescope with the best angular resolution for observing the morphology of these galaxies from $z = 2.5–7$ and the highest sensitivity for these galaxies in the redshift range 2.5–4, where the CDM large scale structure is becoming stronger.

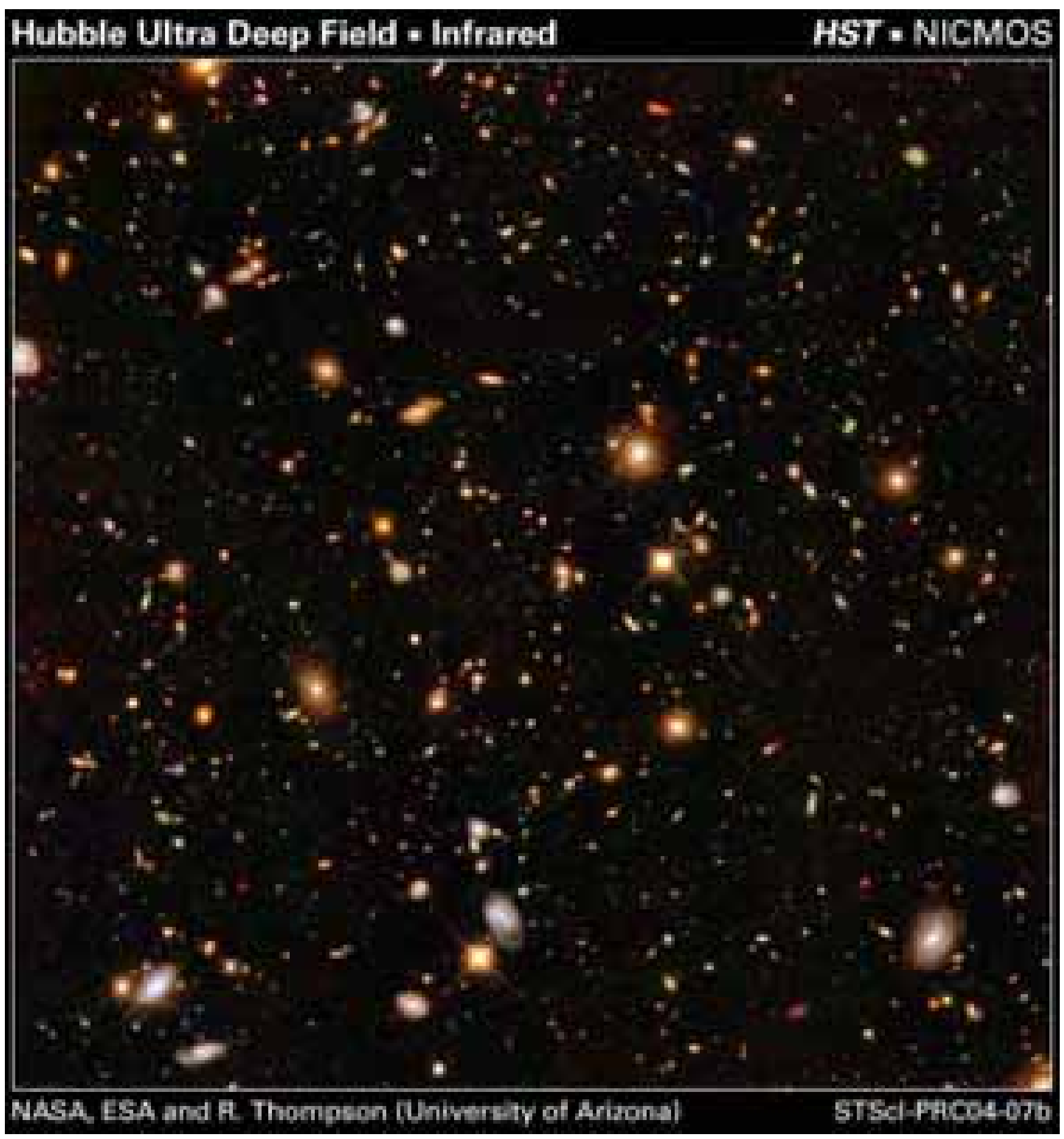

**Figure 10.** The Hubble Ultra Deep Field. A wide-field camera on TPF-C would collect deep fields covering 600 times as much sky, with 6 times the sensitivity and more than double the angular resolution—at the same time as the search for terrestrial planets.

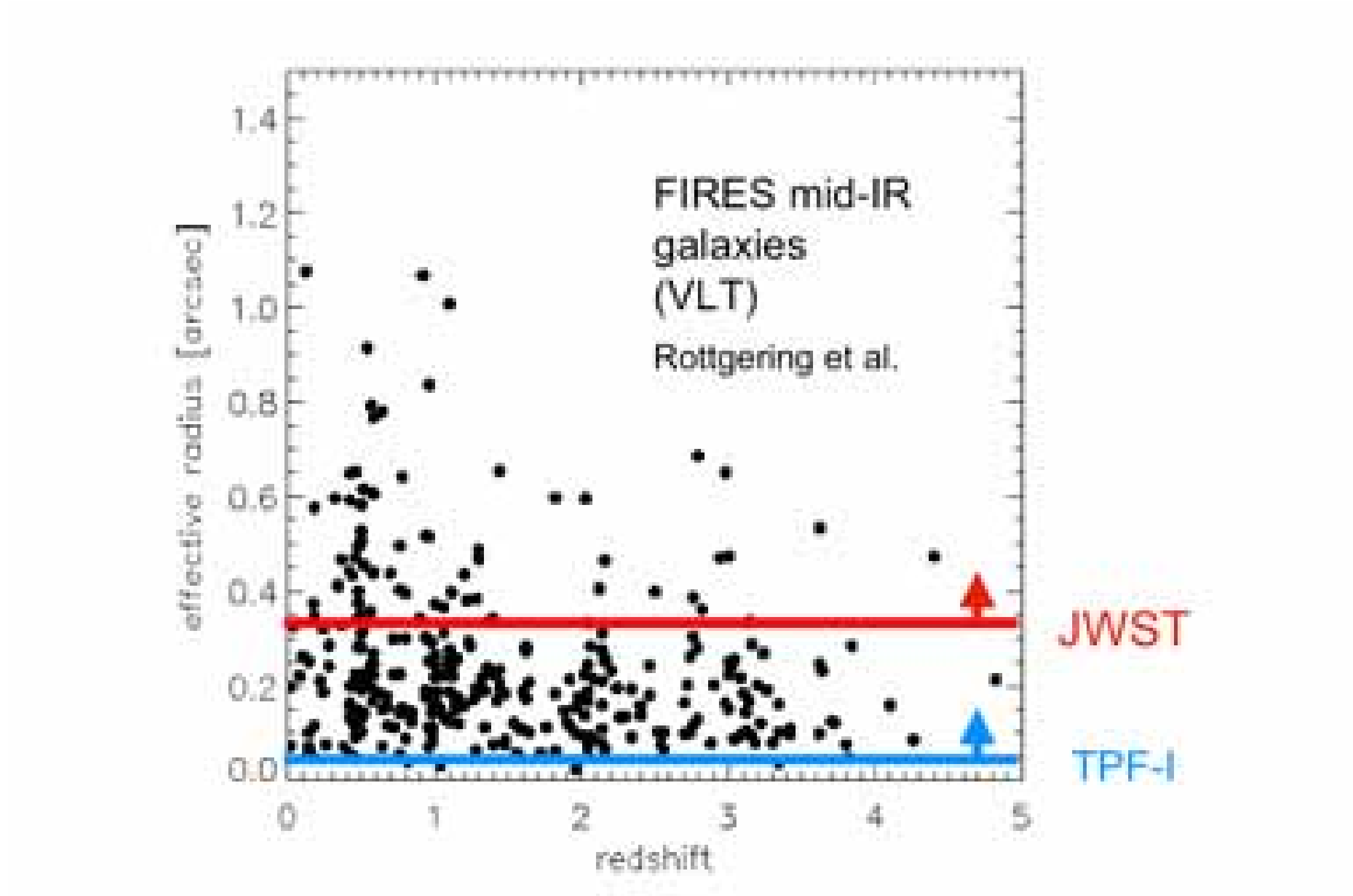

**Figure 11.** Effective radii measured in the K-band as a function of redshift for a subsample of FIRES survey with KAB > 25. TPF-I can resolve most of these galaxies, many more than JWST.

An ideal instrument for examining individual emission line galaxy structures would be an integral field spectrograph in non-coronagraphic mode (pointing off the coronagraphic occulting spot). It would be useful at R~70, but at R~500 it would be more sensitive for detecting emission lines at a given redshift, and at R~1500 it could resolve strong galactic winds. Velocity structures and emission line ratios measured this way would indicate the dominant emission mechanisms, whether they were star formation, AGN, galaxy winds, or shocks. A TPF-C survey of 50 emission line galaxies would require 6 days.

### 5.1.3 Galaxy Morphology at z = 3–5

*Huub Röttgering*

To understand galaxy formation, we would like to start with the physical conditions as they existed in the very early universe, some assumptions about the nature of the dark matter and a suitable choice of cosmological parameters and produce a model that shows how these conditions necessarily yielded the distribution of galaxies and dark matter and black holes in today's universe (e.g., Moustakas and Sommerville 2002; Bell et al. 2003). Models like these often fold in many assumptions about star formation, e.g., the rate at which supernovae and stellar winds which blow gas out of the centers of dark matter halos, the consequence of mergers on star formation, the initial mass function. etc., and attempt to calculate the gross properties of the general galaxy population: the luminosity-redshift-size distribution, the relative numbers of ellipticals and spirals, star formation histories, etc.

TPF-I can provide an essential constraint on these galaxy formation models: the morphologies of very distant galaxies at 1–2 μm in the galaxy rest frame. This band contains the peak of the spectral energy distributions of nearby galaxies, mostly light from older stellar populations. For z ~5 galaxies, this region is redshifted into the TPF-I spectral window.

To illustrate the power of TPF-I for resolving these distant galaxies, consider the Faint InfraRed Extragalactic Survey (FIRES, Franx et al. 2000), a very deep infrared survey centered on the Hubble Deep Field South using the ISAAC instrument mounted on the VLT (Moorwood 1997). Figure 11 shows the effective radius of the objects in FIRES sample as a function of photometric redshift. The angular sizes tend to decrease with redshift. Careful modeling of the various selection effects involved shows that the physical sizes of luminous galaxies ($L_V > 2 \times 10^{10} L_{Sun}$) at $2 < z < 3$ are 3 times smaller than that of equally luminous galaxies today (Trujillo et al. 2003). The size distribution shows that twice the median effective radius is similar to the resolution (FWHM) of the JWST; JWST will hardly resolve these distant galaxies. However, TPF-I can resolve most of the galaxies in this sample, enabling a deep study of the morphologies of older galactic stellar populations at high redshift (Röttgering et al. 2000).

## 5.2 Active Galactic Nuclei

### 5.2.1 Quasar Host Galaxies and the Growth of Supermassive Black Holes

*Michael Strauss and Bill Sparks*

One of the dramatic discoveries in extragalactic astronomy in the last decade is the ubiquity of supermassive black holes in the cores of normal galaxies. Studies of the demographics of galaxies and quasars are consistent with every bulge-dominated galaxy's hosting a supermassive black hole that grew during a luminous phase manifested as a quasar. Nevertheless, this connection is largely statistical; how black holes grow in mass is poorly understood.

By studying the properties of host galaxies of quasars, TPF can test this picture, and investigate the origin of these supermassive black holes. TPF can measure:

1. The distribution and dynamics of gas and stars near the central engines of AGN. Can we see stellar cusps near galaxy centers where stars are pulled into central black holes? Are there small-scale bars which feed the central engine? Are there small-scale gas flows in or out of the AGN? Is there evidence for dust extinction on very small scales? Is there polarized emission on small scales?
2. The morphology, luminosity, and potential well depth (velocity dispersion) of the bulges of the host galaxies. Does the bulge velocity dispersion/black hole mass correlation seen in quiescent galaxies hold for objects accreting near the Eddington limit?
3. The overall photometry and morphology of the host galaxy. Do quasars tend to live in galaxies undergoing interactions? Are there correlations (as hinted at lower luminosity) between morphology and quasar type?

All of the above as a function of redshift, luminosity, and quasar type. Is the feeding of quasars any different at high redshift, when the IGM was much less clustered, but higher density?

TPF-C is ideally suited to tackle this range of questions—requiring observations in the visible and near-IR of low-surface-brightness extended emission within a few arcseconds of a very bright point source. Since a kiloparsec subtends of order one arcsecond at cosmological distances, TPF-C may be able to resolve distances as small as 50 pc, and even smaller than that for very nearby quasars (e.g., 3C 273 at $z = 0.158$). No other planned mission can approach this angular resolution at high dynamic range.

This science requires direct pointings to known quasars and doesn't benefit from parallel observations during the planet search. However, a typical quasar host galaxy subtends ~10 kpc, or 10" at the most, and most of the action (see the list above) is on sub-galactic scales, so a large field of view is not needed. The desired new capabilities are narrow-band imaging or an IFU (R = 1000–3000) to study stellar dynamics and line emission in outflowing and infalling gas, and imaging polarimetry or imaging spectropolarimetry to look for synchrotron emission from jets, and study scattering.

### 5.2.2 Galaxies and the Intergalactic Medium Along the Lines-of-Sight to Quasars

*Joop Schaye*

With its ability to detect faint galaxies near much brighter quasars, TPF-C can probe the interactions between distant galaxies and their environments and provide a means to study galaxies selected in absorption, offering us a relatively unbiased view of the universe.

Feedback processes from star formation are thought to drive supersonic outflows from galaxies into intergalactic space. These galactic winds shock-heat the environments of galaxies and pollute them with heavy elements. Such feedback processes are thought to have a profound impact on the galaxies themselves as well as on subsequent generations of galaxies.

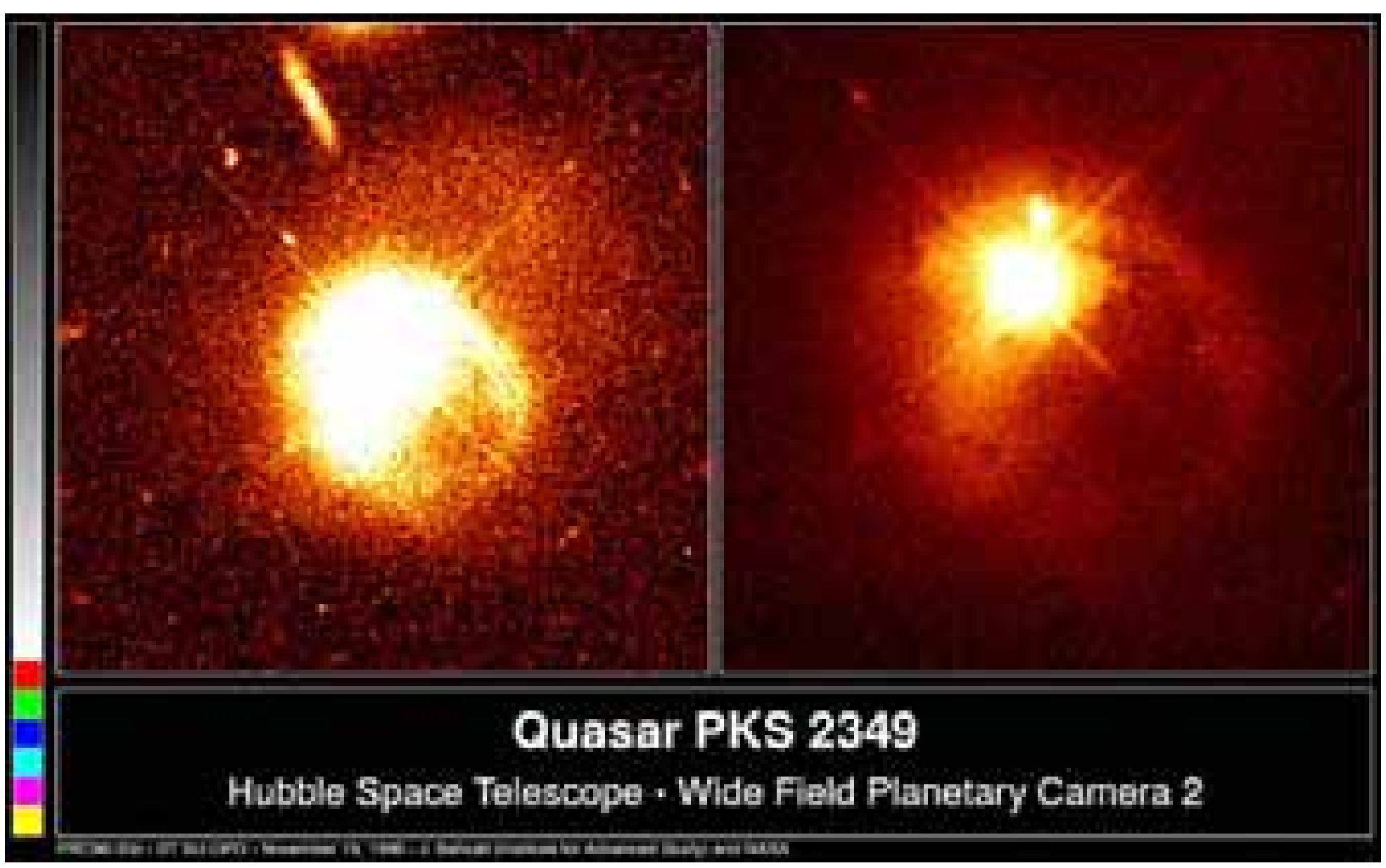

**Figure 12.** HST image of quasar PKS 2349 and its host galaxy, apparently distorted by the gravitation of a central supermassive black hole (Bahcall et al. 1995). The left image is a re-binning of the right image to bring out faint features. Images of these sources with higher angular resolution and higher contrast with TPF-C can show how AGN form.

Quasar absorption line spectroscopy reveals galactic environments that would be too dilute to observe in emission—but it does not provide much information about the galaxies themselves. TPF-C can directly detect the galaxies whose environments are revealed in the quasar spectra, allowing us to interpret the information encoded there. Intensive observing campaigns with 8-m-class telescopes (e.g., Adelberger et al. 2003) have proven the usefulness of such observations, but these campaigns are severely limited in their ability to observe galaxies near the line-of-sight to the quasar due to the large contrast in brightness between the quasars and the galaxies. The galaxies closest to the line-of-sight are precisely the ones expected to have the greatest effect on the gas seen in absorption.

Quasar absorption lines also provide us with a means to study the chemical evolution and other properties of distant galaxies in a way that is much less biased towards high-density regions of the universe than studies of galaxies selected in emission. Relating the absorption line studies to observations of galaxies in emission requires establishing the nature of the galaxies that are giving rise to the absorption: faint galaxies near the line-of-sight to a background quasar.

The high quality quasar spectra needed for these projects are presently only available for a few tens of quasars—though future ground-based telescopes could provide more. The wavelengths of these high-quality absorption line spectra (0.3–0.8 μm) correspond to rest-frame UV (912–1600 Å) in intervening galaxies and IGM at redshifts of roughly 2–4. The target galaxies will be 1–2 arcseconds or less from the quasar at most; ideally one would like to detect the emission from the same gas that is doing the absorption. The galaxies are typically 5 or more magnitudes fainter than the quasars.

Relating galaxies to quasar absorption features requires measuring the galaxy redshifts. The more precise the measurement, the better, but R = 300 will suffice for a study of galactic properties, since it will be rare to have several galaxies close to the line of sight; it should be obvious which galaxy is responsible for the strongest absorption lines. Many models predict that the typical absorber will be 0.1 $L_*$, which corresponds to an R band magnitude of 27.4 for a galaxy at z = 3.

Using quasars to investigate interactions between galaxies and their environments requires higher spectral resolution (R = 1000–3000) to measure redshifts accurately enough to locate the galaxies to a Mpc or better. Increasing the resolution beyond 100 km/s (R = 3000) probably would not help this effort because of redshift space distortions. Adelberger et al. (2003) measured R~600 spectra of galaxies brighter than R = 25.

### 5.2.3 AGN Tori

*Huub Röttgering and Bill Sparks*

In the currently popular and attractive "unified" model of Active Galactic Nuclei (AGN), all AGN contain a central black hole fed by an accretion disk surrounded by an optically thick obscuring torus. The orientation with respect to the line of sight determines whether the object is observed to have broad emission lines, originating from within the hot central hole of the torus, or whether the torus blocks this region from view, leaving only the unobscured narrow line region visible.

Although this picture is capable of explaining a large number of the differences among the various classes of AGN (e.g. Antonucci 1993, Urry & Padovani 1995), a vigorous debate remains whether other mechanisms contribute to, or even dominate, AGN diversity. For example, it has been argued that in a subset of AGN, the main power-source is not the black hole but supernova explosions produced within a central starburst region. It is not clear, however, if all AGN have starburst regions, whether all starburst galaxies contain AGN, or what the causal relation is between these two phenomena.

If the AGN unification theory is right, a free-flying TPF-I will be able to produce the first detailed maps of nearby AGN tori greatly surpassing the capabilities of ground-based interferometers. Recently the VLT interferometer carried out the first mid-infrared *interferometric* spectral observations of the nearby AGN, NGC 1068, at a resolution of ~30 milliarcseconds (Jaffe et al. 2004). These observations seem to indicate a compact dust structure only a few parsecs across around an unresolved hot core, possibly the accretion disk. They also suggest dust of distinctly different composition from that common in our Galaxy. The VLT-I can map only the few tori that are brighter than 1 Jy, and eventually only provide ~20 visibility points. However, TPF-I can provide two or three orders of magnitude more visibility points and μJy sensitivity, allowing us to study the dynamics of these objects and their evolution and classification.

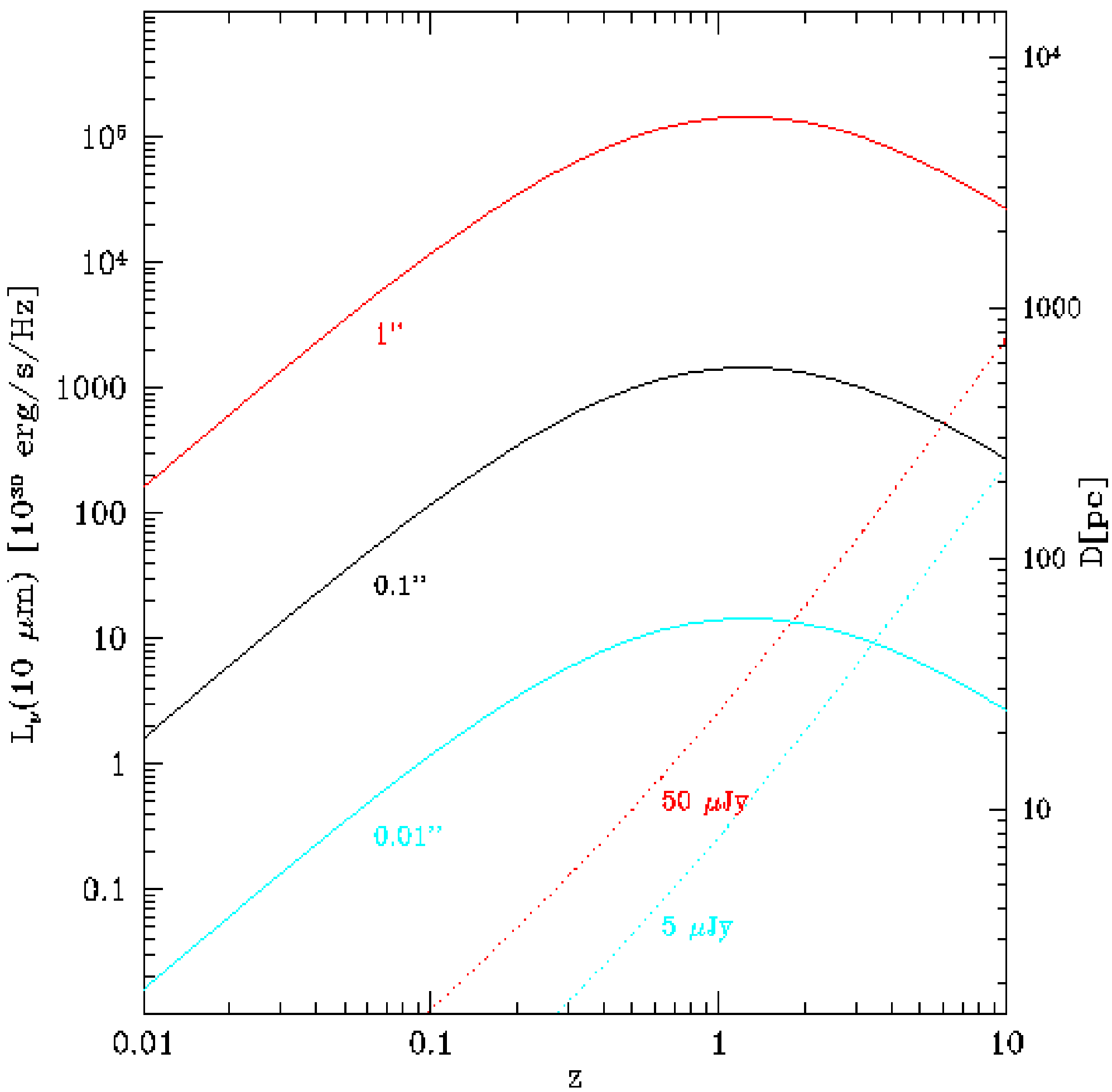

**Figure 13.** Luminosity as a function of redshift for AGN Tori with three different angular sizes according to the model of Granato et al. 1997 (solid lines). Dotted lines show detection limits for 50 µJy sensitivity and for 5 µJy sensitivity. These models suggest that TPF-I can detect and resolve AGN at the large end of the luminosity/size range at redshifts z > 10.

A relatively weak AGN like NGC 1068 has a 10 µm luminosity of ~$1.7 \times 10^{31}$ erg $s^{-1}$ $Hz^{-1}$ and a 60 pc torus according to the models of Granato et al. (1997). Up to redshifts of z = 1–2, such weak AGN are brighter than the nominal flux needed for TPF-I to produce complex images (e.g. Röttgering et al. 2000). But TPF-I can also study dusty tori at large redshifts. Assuming the Granato et al. (1997) models with torus diameters of 300 times the dust sublimation radius and nominal TPF-I resolution of 20 mas at 10 µm, brighter AGN can be mapped up to redshift of z = 10–20 if they exist at these redshifts. These high-z observations will show how the properties of tori change with redshift and when and how these tori and their associated massive black holes form.

# 6 Cosmology

## 6.1 Dark Energy

*David Spergel and Mustapha Ishak-Boushaki*

Dark energy is one of the most profound mysteries in science. We can only directly measure a few of the basic properties of the dark energy: its clustering properties (usually characterized by a sound speed, $c_s$) and the time evolution of its energy density, $\rho(a)$. The evolution of the energy density is often parameterized as $\rho(a) = \rho_0 a^{-(1+3w(a))}$. If the dark energy density is a cosmological constant, then $w(a) = -1$ and the dark energy makes a significant (> 25%) contribution to the total energy density of the universe only for $z < 1.0$. On the other hand, if the dark energy density/matter density is a slowly varying function of time, then the dark energy could make a significant contribution to the energy density of the universe even at high redshift.

Type Ia supernovae are a powerful tool for probing the dark energy. When properly calibrated, they are standard candles that can by used to measure the luminosity distance/redshift relation. Over the past several years, ground-based surveys using ~100 supernova, primarily at $z < 0.8$, have provided direct evidence for the acceleration of the universe. A recent sample of 7 high redshift supernovae ($z > 1.2$) that have been detected with HST show clear evidence for deceleration at high redshift. The addition of this relatively small sample of high redshift supernova not only provides important supporting evidence for the existence of dark energy, but also places powerful constraints on its properties. The HST sample improves the limits on dw/dz by a factor of 8!

Over the coming decade, several different ground-based surveys will increase the sample of low redshift ($z < 1$) supernova by roughly an order of magnitude. The on-going ESSENCE survey should detect ~400 new supernova. The planned LSST survey should collect nearly 4000 supernova. If the dark energy is a cosmological constant, then these low-redshift samples should be able to place strong constraints on w. On the other hand, if dark energy plays an important role earlier in the history of the universe, then a large sample of high redshift supernova are needed to measure the time evolution of the dark energy.

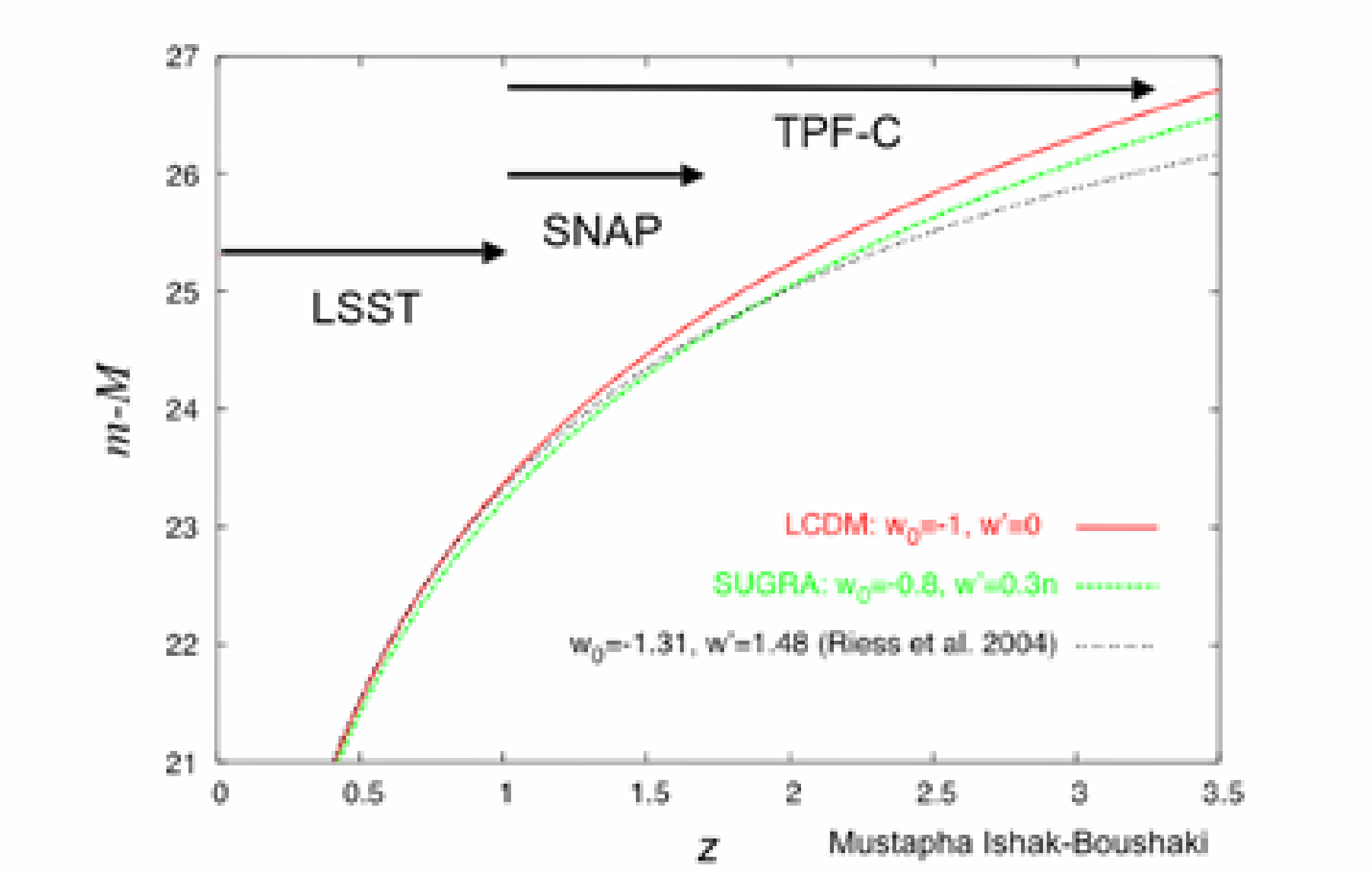

**Figure 14.** Luminosity distance as a function of redshift for three different cosmological models, showing how supernovae at z=2-3 accessible to a wide field camera on TPF-C will have the most leverage for distinguishing one model from another (Mustapha Ishak Boushaki).

If TPF-C is equipped with a wide-field camera that extends into the near infrared, then it should be able to detect a very large sample of high redshift supernovae. With an undersampled 5' × 5' CMOS detector operating out to 1.7 μm, TPF-C while in parallel mode will survey more than 10 times the area of GOODS survey and roughly 8 times the integration time. With 2000-second integrations, the HST survey was limited to detecting supernovae with $z < 1.6$. With TPF-C's sixfold increase in collecting area and with much longer integrations on each field, it should easily be able to detect supernovae out to $z \sim 3$. Beyond this redshift, K-corrections are large for SNIa's. Scaling from the HST GOODS discovery rate, TPF-C should be able to detect over 1000 supernovae. Unlike the SN sample measured by the proposed SNAP telescope, almost all of the TPF-C supernovae would have high redshifts.

The TPF-C supernova sample is an ideal complement to the proposed LSST sample. Figure 14 shows how a TPF-C supernova survey can distinguish between dark energy models that are almost indistinguishable in lower-redshift experiments. The curves in the figure correspond to the standard Lambda-Cold-Mark-Matter model (LCDM) with the Einstein cosmological constant, a scalar field dark energy model motivated by super-gravity theory (SUGRA), and a quintessence scalar field model that is the best fit to the recent supernova analysis from Riess et al. (2004). Only within the high redshift range probed by TPF-C are all three dark energy models clearly distinguished.

TPF-C can perform follow-up spectroscopy on the detected supernovae with the IFU in the coronagraph. Since supernova spectroscopy is limited by the surface brightness of the host galaxy, TPF-C's high spatial resolution will help it quickly obtain spectra of high redshift supernovae. This follow-up program would be similar to the piggyback follow-up observations that used HST STIS to study the supernova detected in the GOODS fields.

TPF-C's deep optical and near-infrared capability would be unique. At 0.4–1 μm, TPF-C has much higher resolution than any ground-based telescope or JWST. Ground-based telescopes are severely limited by the atmosphere at 1–1.7 μm. Although as a cryogenic telescope, JWST's infrared sensitivity should surpass TPF-C's, it is unlikely that NGST will devote 4 years of observing time to deeply imaging of a handful of fields at a cadence of roughly once every 1–3 months. For TPF-C, the deep imaging and variability search program can be done in parallel to its program of planet search and characterization. The TPF-C variability search will complement the planned LSST program. LSST will detect variable objects (V = 26) over 3π steradians. TPF-C will be able to detect much fainter objects (V = 33) over a much smaller portion of the sky (roughly 1 square degree).

## 6.2 Dark Matter

*Simon DeDeo*

Recent HST ACS observations of the rich cluster Abell 1689 contain a stunning number of multiply imaged background galaxies; at least thirty image sets have been identified, containing more than a hundred images and a number of new giant arcs where background galaxies have happened to lie near caustics. This kind of strong lensing has long been used as a probe of the dark matter distribution in clusters; the positions and shapes of the lensed images indicate the dark-matter mass distribution, which can be reconstructed via models. Usually, the prominent giant arcs in an image provide the greatest amount of information. The nature of dark matter is a great unanswered question in cosmology, and the distribution of the dark matter is one of the few handles we have on its nature.

Figure 15 shows how TPF-C may lead to a qualitative—not just quantitative—improvement in our understanding of dark matter from such observations of rich clusters. The figure shows a UV GALEX image of the Whirlpool Galaxy, M51, lensed by a simulated cluster mass distribution from an N-body simulation, as seen by HST's ACS camera, and a wide-field camera on TPF-C. The simulated images of this distorted giant arc similar to those seen in Abell 1689 include sky background and photon noise, assuming reasonable exposure times and noise levels.

Unlike HST, TPF-C will be able to see much more than just the general shape of the giant arc. The simulated HST image shows some substructure, and the HST image of Abell 1698 shows a few prominent HII regions, but HST has little ability to measure the morphology of HII regions. TPF-C can clearly resolve the HII regions in the lensed galaxy.

Comparing the shapes of the same HII region in different TPF-C images will reveal the phenomenon of microlensing by mass fluctuations in the cluster matter distribution down to a level of approximately $10^6$ $M_{Sun}$. Because of the paucity of detail in the HST images and because their shapes depend in a highly non-linear fashion on the matter distribution, the matter distribution in a cluster must be parametrized, i.e., the images are used to constrain models, not to derive the mass distribution ab initio. However, TPF-C's large aperture and greater resolution should yield at least four times the number of images found by HST in a cluster such as Abell 1689; each resolved HII region can be considered a separate image for the purposes of the reconstruction. This abundance of information may permit a completely non-parametric reconstruction of the dark matter distribution or at least greatly expand the possible complexity of a parametrized matter distribution model. The many images of lensed HII regions will allow us to study a cluster's dark matter structure on scales ranging over a range of perhaps a factor of a thousand, from Mpc down to kpc, in a single observation, probably an unreachable goal by any other means.

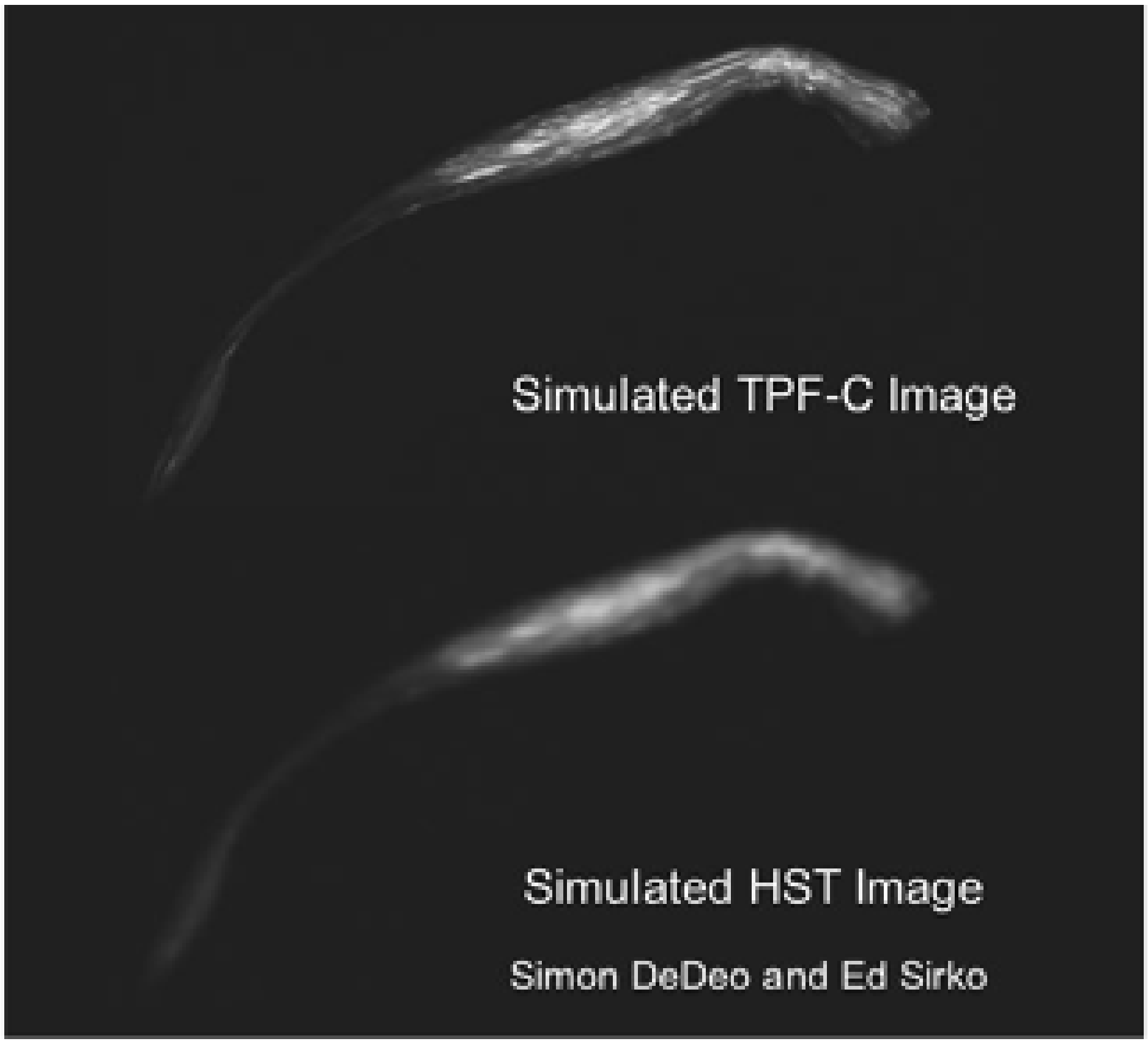

**Figure 15.** Simulated images of an arc lensed by a clumpy dark matter halo from the HST and from a wide-field camera on TPF-C. Because it can resolve HII regions in the lensed galaxy, the TPF-C image provides at least 4 times as many constraints on the lensing halo's mass distribution (Simon Dedeo and Ed Sirko).

The improved resolution of TPF will also allow better identification of image sets; because the images in the field will be both faint and numerous, spectroscopic identification is impractical and flawed because of microlensing. Instead, the simplest method of identifying which images came from the same galaxy is through an iterative model beginning with "obvious" pairs. TPF-C's ability to resolve substructure in images will greatly simplify this task, because image sets can be identified on the basis of internal structure.

# 7 Capability Enhancements for the TPF Coronagraph

## 7.1 Wide-Field Camera

*Doug Lisman and Tony Hull*

A wide-field camera can work in parallel with the coronagraph camera, sharing the focal plane. Such a device could produce something like a Hubble-Deep-Field image during every planet-search integration—with higher angular resolution, sensitivity, and potentially, with a larger field of view. A field of view (FOV) > 3 arc minutes, would include sufficient background objects to enable astrometry, given a well-sampled PSF. The estimated accuracy would be 100 μ arcsec for objects as faint as V = 19–23. The baseline TPF-C design includes a placeholder Wide Field Camera (WFC) concept, which we will discuss here.

The baseline TPF-C telescope is an off-axis Cassegrain with 8m × 3.5m elliptical aperture and 140-m effective focal length. A fold mirror bends the beam behind the primary mirror where a fixed mirror at the Cassegrain focus picks off light to the WFC, and a central hole in the pickoff mirror passes light on to the starlight suppression system. The WFC beam passes through structure at a distance of 2m from the pickoff mirror; this structure restricts the beam diameter to ≤ 30 cm at this point. The WFC is located slightly beyond this restricting structure where a larger volume is available. The focal plane could contain, e.g., 8×8 butted CCD detector arrays (4096×4096 13.5μm pixels each) for a visible-light camera, or 8×8 butted HgCdTe arrays (2048×2048 27μm pixels each) for a near-IR camera.

Equation 1 gives an estimate of the FOV as a function of field size constraint (H), distance of the constraint from the pickoff mirror (d), the telescope maximum dimension (D), and focal length (f). Figure 16 shows that our placeholder concept supports a FOV slightly greater than 4 arc-minutes.

(1) $$FOV_{rad} = \frac{H - dD/f}{f + d} \qquad \text{[rad, m]}$$

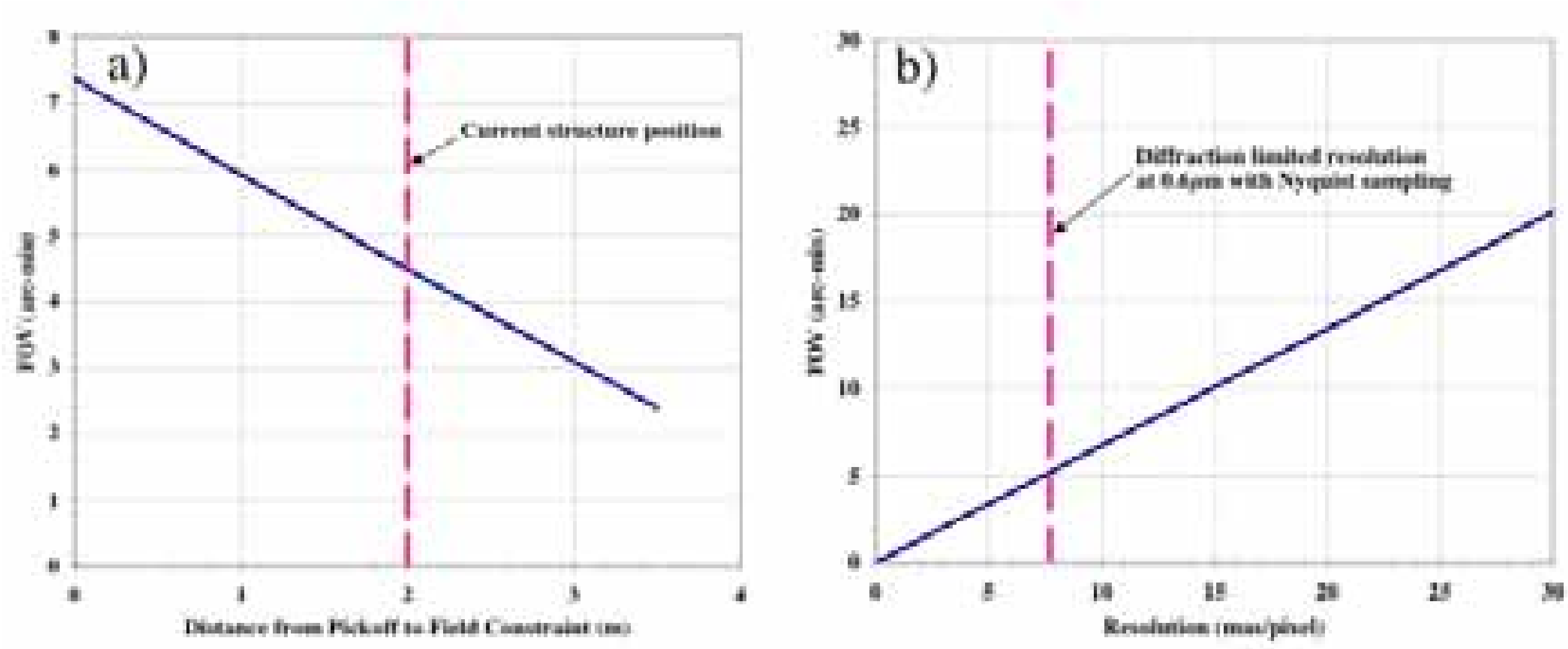

**Figure 16.** Two constraints may limit the field-of-view to 4–5 arcminutes: (a) the larger the distance from the pick-off mirror to the field size constraint, the smaller the field-of-view; and (b) the higher the required angular resolution, the smaller the field-of-view.

Unfortunately, working in an L2 or drift-away orbit will place large constraints on the data rate, and hence the useful FOV. A visible camera with a 5’ × 5’ FOV would produce ~$10^{12}$ bits per day, a 1.0–1.7 μm camera $10^{11}$. This situation can be addressed with data compression, windowing, and on-board data processing, and by upgrading the deep space network.

.

Here is a bounding analysis for comparison with the beam clearance constraint discussed above. We assume a downlink data rate capability of 64 Mbps, consistent with the needs of JWST that are currently driving an upgrade to DSN capabilities. Equation 2 gives an estimate of the required downlink data rate as a function of FOV and resolution. For the parameters shown, this reduces to the relationship: FOV = 0.67*Resolution [rad, mas/pixel], assuming continuous operation and no data compression. Figure 16 (b) shows that even with these conservative assumptions, the FOV capability at diffraction-limited resolution is less constrained by the downlink capability than the field size constraint discussed above.

$$(2) \quad Data_Rate = \left(60{,}000\frac{FOV_{arc-\min}}{R_{mas/pixel}}\right)^2 \times \frac{t_{exposure}}{t_{read}} \times \frac{K_{bits/pixel}}{t_{downlink}} < 64\times 10^6 bps$$

Where: K = 6.6 bits/pixel, for the product of:

- Telemetry bits per pixel = 12
- Minor axis resampling factor = 0.5
- Telemetry overhead factor = 1.1
- Data compression factor = 1

$t_{exposure}$ = average daily total exposure time = 24 hr

$t_{downlink}$ = daily downlink duration = 4 hr

$t_{read}$ = period between reading out the detector = 1,000 s

Caveats:

1) TPF-C is optimized for guiding on an exoplanet target, with a bright star V < 6 exactly at the center of the field. Guiding without a bright central star would be much less accurate (~1 arcsecond), unless auxiliary guiding instrumentation were added.
2) Cooling of a large detector within a highly stable thermal enclosure may be a challenge.

## 7.2 UV Camera

*Chuck Bowers and Bruce Woodgate*

If some of TPF-C's mirrors were coated with aluminum instead of silver, it could conceivably support a UV camera operating down to 200 nm. The primary coronagraph probably requires mostly silver mirrors for acceptable throughput in the 0.5–0.8 μm band. However, if the primary and secondary mirrors of TPF-C were aluminum, and the rest silver, the net reflectance would be acceptable at 0.5–0.8 μm. A pickoff mirror following the telescope could then feed a separate UV camera operating down to 200 nm. Figure 17 illustrates this point showing the relative system reflectance of a configuration including two Al and 18 Ag mirrors; for packaging purposes, it might be necessary to coat a tertiary mirror as well with Al.

One complication is that both Ag and Al mirrors require protective overcoatings; Ag needs protection from tarnish, Al from oxidation. These overcoatings alter the polarization properties of the mirrors. Polarization adds low-order aberrations that reduce the coronagraphic contrast of TPF-C, so it is important to minimize the polarization induced in the reflected beam, often parameterized in terms of the phase difference between the components of the electric field parallel and perpendicular to the mirror's grade.

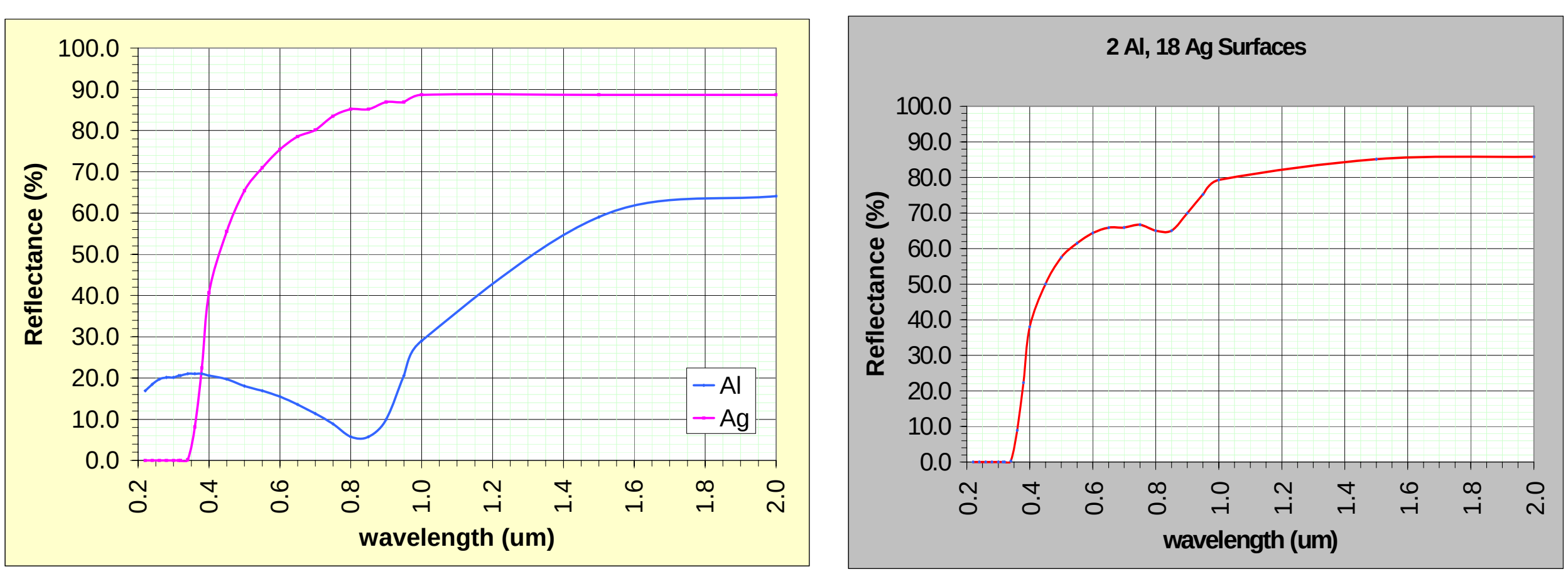

**Figure 17.** Net reflectance for 20 surfaces of aluminum and silver (left) showing why a TPF-C planet search at 0.5-0.8 μm requires silver coatings. But if the primary and secondary mirrors were aluminum, and the rest silver, the net reflectance would be acceptable at 0.5-0.8 μm (right), and TPF-C could have a UV camera operating down to 200 nm.

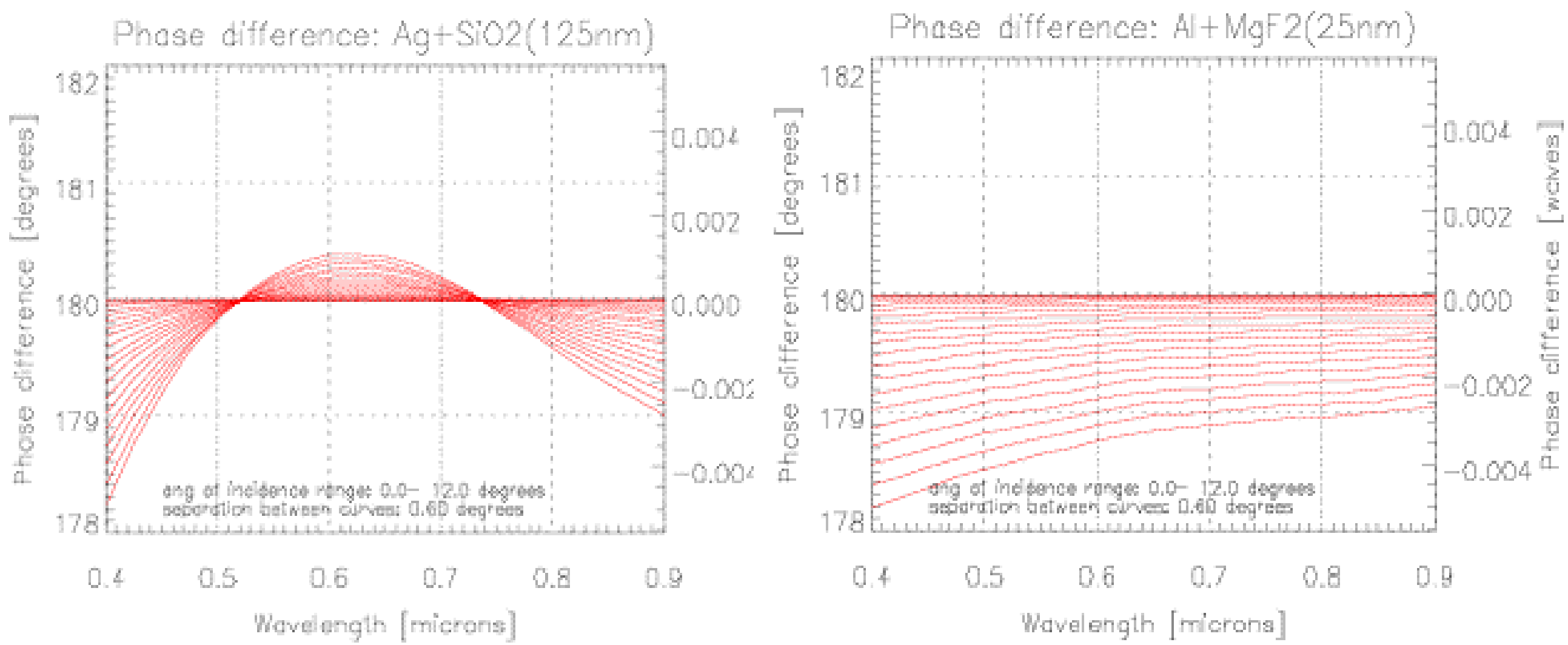

**Figure 18.** The phase difference between components of light parallel and perpendicular to the grade reflected from a possible TPF-C mirror as a function of wavelength; each red curve shows a different angle of incidence. The silver mirror with an optimized $Si0_2$ coating (left) produces much less phase difference than the aluminum mirror with 25 nm $MgF_2$ coating (right).

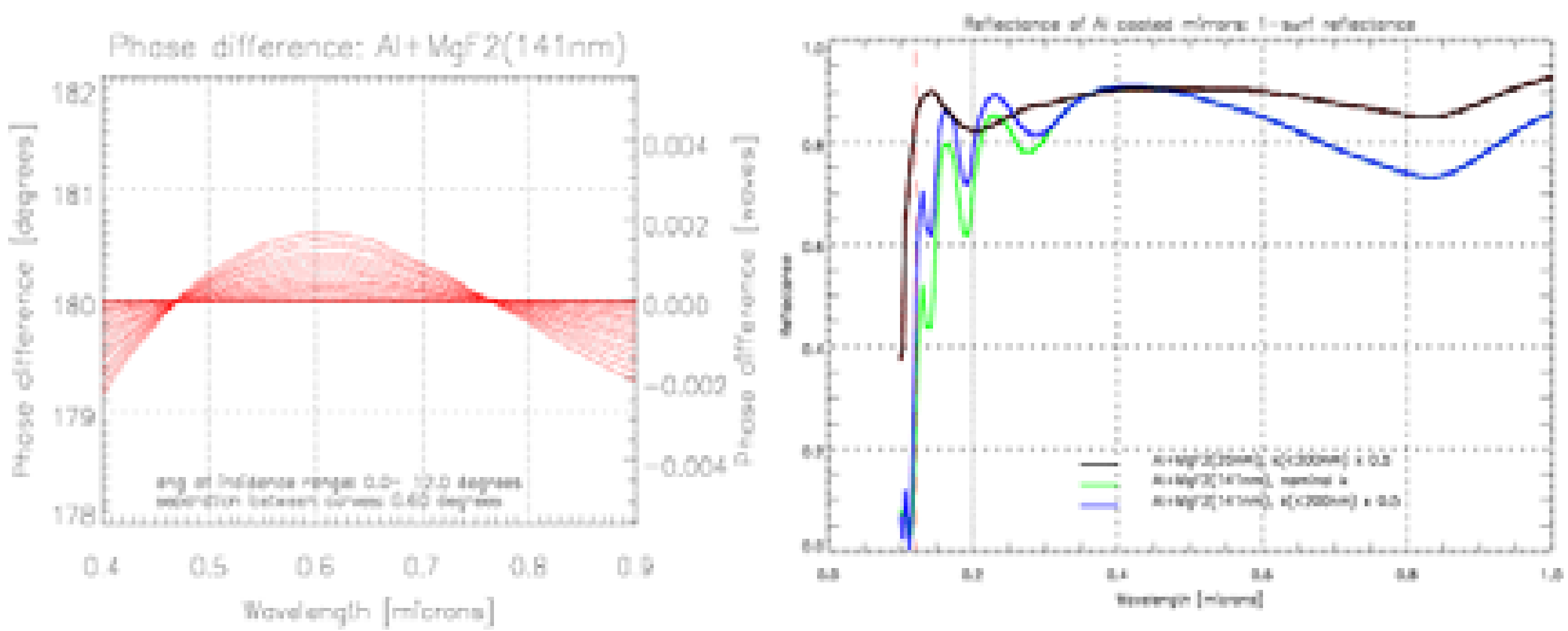

**Figure 19.** Although an $MgF_2$ coating on an Al mirror can be thickened to minimize the phase difference across the TPF-C 0.5–0.8 micron planet-search bandpass (left), the reflectivity of the resulting mirror is likely poor at 0.8 microns.

Figure 18 shows this phase difference for an overcoated Ag mirror whose $SiO_2$ overcoat thickness was adjusted to minimized this phase difference over the TPF-C bandpass, minimizing the induced polarization. The phase difference is plotted for several angles of incidence from 0–12°. It also shows the phase difference for a UV coating of Al+$MgF_2$ (25nm), nominally used to extend high efficiency to wavelengths short at 121.6nm. The superior polarization performance of the overcoated Ag mirror is evident. The lower performance UV-Aluminum coating may still be acceptable for coronagraphy but requires further detailed analysis.

Alternatively, the thickness of the $MgF_2$ overcoat on Aluminum may be increased to produce a coating that induces a minimum degree of polarization in the 0.5–0.8 µm band. A layer about 140 nm thick produces good results as shown below for a single mirror coating; TPF-C altered for UV camera use would require at least 2–3 such mirrors. While this coating has improved visible polarization performance, its reflectivity in the ultraviolet below 200nm is significantly reduced by $MgF_2$ absorption and modulated by interference within the film. This reduction is illustrated in Figure 19 using two models of ultraviolet $MgF_2$ absorption. In addition to the reduced ultraviolet reflectance, the thickness of this overcoat produces an interference minimum nearly coincident with the intrinsic Al dip near 800 nm, lowering the visible performance of such a coating as well.

# 7.3 Expanded IFU

*Bruce Woodgate*

The base TPF-C design will likely use an integral field unit (IFU) for spectral deconvolution and spectroscopy. However, an additional IFU with higher spectral resolution or larger field of view could be a powerful multi-purpose instrument for general astrophysics applications including circumstellar disks, supernovae, QSOs, stellar populations, and galactic dynamics. An integral field spectrograph (IFS) would sit behind the pupil of the coronagraph. It would require a large detector array, with tens of Mpixels. This array could be shared with a wide-field camera. The table below shows a few suggested IFU designs. The first (Mode 1) is one designed for the terrestrial planet search. The others offer various combinations of spatial and spectral coverage for general astrophysics.

Caveats:

1) Detector noise will limit TPF's spectrographic sensitivity for long exposures if it uses the best current imaging detectors (CCDs with QE ~80% and read noise ~2 els); zero read noise (photon counting) detectors are needed. Even at ~40% quantum efficiency, these devices are better than the CCDs for exposure times of > 2–3 hours.

2) Like the wide-field camera, expanded IFUs would generate a high data rate.

**Table 1.** Suggested Integral Field Unit Designs for TPF-C

| IFU Mode | 1 | 2 | 3 | 4 |
|---|---|---|---|---|
| R | 70 | 3000–20,000 | 3000 | 3000 |
| Pix/spectrum | 100 | 100 | 4000 | 4000 |
| Angular resolution per pixel | 37 mas | 37 mas | 37 mas | 220 mas |
| FOV | 3.7×3.7” | 3.7×3.7” | 3.7×0.185” | 3.7× 3.7” |
| Disperser | Prism | Grating | Grating | Grating |

# 8 Capability Enhancements for the TPF Interferometer

## 8.1 High-Resolution Spectroscopy

The mid-IR is rich in spectral features from source components in the solid state and gas phase. The spectrum includes atomic hydrogen recombination lines, ionized Ne, Ar, and S forbidden lines, $H_2$ rotational lines, and silicate, SiC and polycyclic aromatic hydrocarbon (PAH) features. These features appear on a continuum of thermal dust emission. A wealth of information is available from the mid-IR spectrum, such as the radiation field intensity and hardness, the dust temperature, and the chemical state of the medium. Kinematic information is potentially available as well, but only when the available spectral resolution is ~$10^4$ for galaxies or ~$10^{5\text{-}6}$ for protostars and objects of similar size.

The spectral resolution of TPF-I can be increased by increasing the stroke of the delay line. R = 1000 at 10 μm requires only a 1 cm stroke, but R = $10^6$ requires a 10-m delay line stroke, a packaging challenge. Such very high resolutions probably require the addition of a grating.

## 8.2 Double Fourier Interferometry

*David Leisawitz*

The wide field-of-view double Fourier technique (e.g. Mariotti and Ridgway 1988) can enable a TPF mid-IR interferometer to provide high spatial resolution, high spectral resolution observations of spatially extended astrophysical sources. TPF-I could map protostars, debris disks, extragalactic star forming regions, and protogalaxies on relevant spatial scales and simultaneously provide the spectroscopic data that would enable deep insight into the physical conditions in these objects. The table below shows desired measurement capabilities for a variety of targets and indicates that 10 mas is an interesting angular resolution and ~1 arcmin is an interesting FOV for a wide variety of applications.

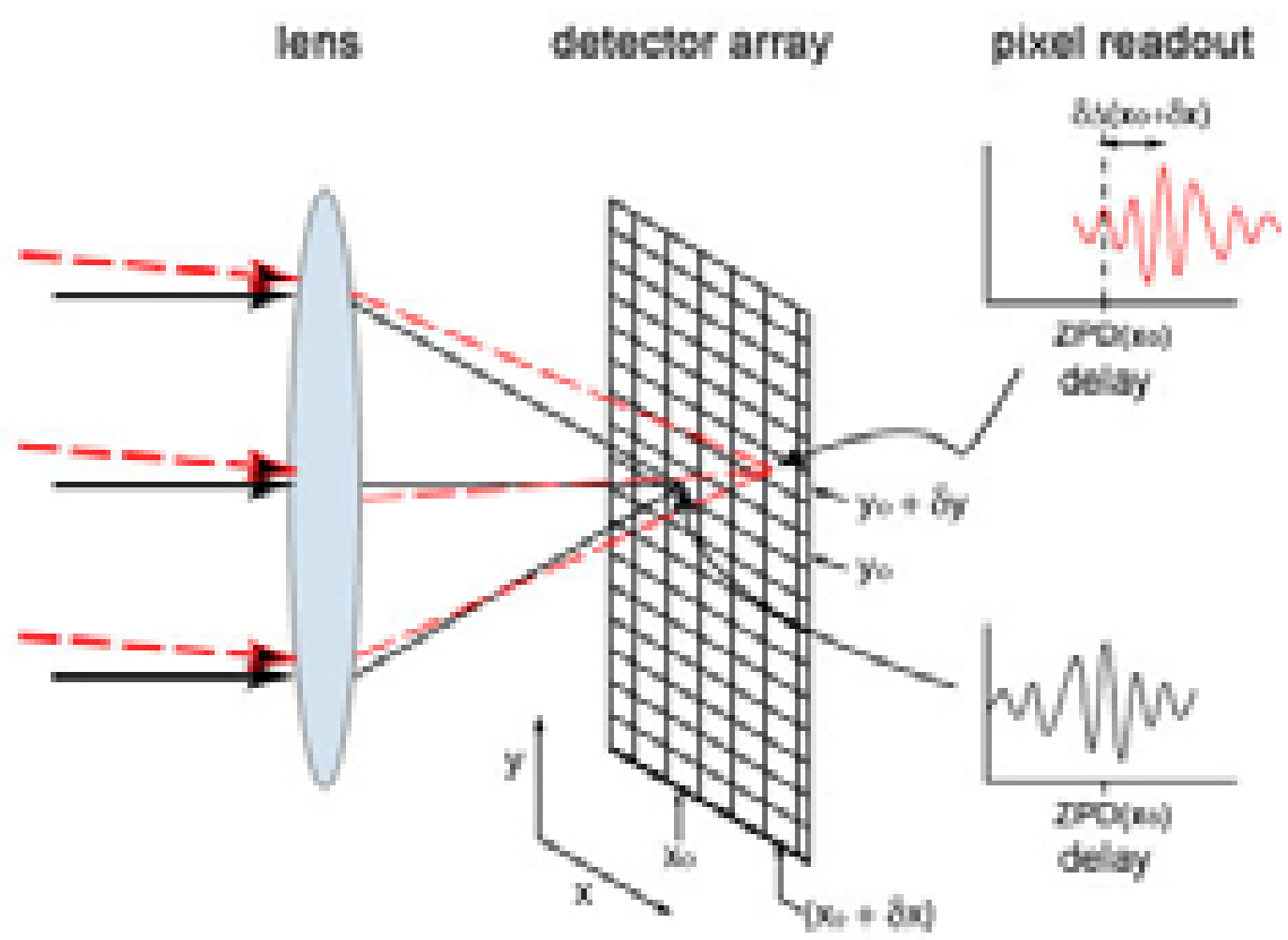

**Figure 20.** In double-Fourier interferometry, a single-pixel detector records one fringe from on axis light (black lines) as the delay line scans. In the Wide-Field Double-Fourier technique, the detector records fringe patterns from many off-axis pointings (red dotted lines) simultaneously as the delay line scans. The off-axis pointings have different zero path difference (ZPD) delays.

The idea behind the double Fourier technique is that a Michelson stellar interferometer equipped with a pupil plane beam combiner and a scanning optical delay line can be operated like a Fourier transform spectrometer (FTS). Instead of providing only a visibility measurement for the interferometer baseline established by the collecting aperture locations, such a device produces an interferogram whose 1-D Fourier transform is the spectrum of the target scene on the spatial scales to which the interferometer is sensitive. Combined, the interferograms from all the baselines provide a three-dimensional data cube where the cube has two spatial and one spectral dimension, like the data from the integral field units discussed above.

Using a conventional double-Fourier system, a TPF interferometer with 4 m diameter collectors operating at $\lambda = 10$ μm with a maximum baseline of 300 m could image a 0.6 arcsec diameter FOV at 4.8 mas spatial resolution. This field of view would be inadequate for the science programs mentioned above. However, the Wide-field Imaging Interferometry Testbed (WIIT) at NASA's Goddard Space Flight Center was designed to develop and demonstrate a technique for wide-field (i.e., FOV >> $1.2\lambda/D$) imaging in which a detector array is used to enhance the spatial multiplexing efficiency (Leisawitz et al. 2003). In this design, light from field angles $\theta >> 1.2\lambda/D$ relative to the principal axis of the interferometer focuses onto additional pixels in a detector array, which record interferograms shifted by a geometric delay corresponding to $|\mathbf{b}|$ times the sine of the component of $\theta$ aligned with the baseline vector **b**. The field of view accessible to an interferometer like WIIT is given by $\theta_{FOV} = N_{pix}\theta_p/2$, where $\theta_p = 1.2\lambda/D$ is the primary beam diameter, $N_{pix}$ is the number of pixels along one dimension of the detector array, and the factor 2 allows for Nyquist sampling of the primary beam. For a $256^2$ pixel array ($N_{pix} = 256$) on the interferometer described above ($D = 4$ m, $\lambda = 10$ μm), $\theta_{FOV} = 79$ arcsec, a good match to the requirements outlined in Table 2.

**Table 2.** Desired measurement capabilities: angular scales

| Ancillary Science Target | Interesting Physical Scales | Typical Distance | Interesting Angular Scales (arcsec) | |
|---|---|---|---|---|
| | | | Desired Resolution | Desired FOV |
| **Protostar** (envelope, disk, outflow) | 1 Š $10^4$ AU | 140 pc | 0.007 | 70 |
| **Debris disk** | 1 Š 300 AU | 3.2 pc (e Eri) | 0.3 | 93 |
| | | 30 pc | 0.03 | 10 |
| **Extragalactic Giant H II Region** | 1 Š 100 pc | 5 Mpc | 0.04 | 4 |
| **Coma cluster galaxy** | 0.01 - 10 kpc | 107 Mpc | 0.02 | 19 |
| **High-z protogalaxy** | 1/100th source to separation between merging systems | N/A | 0.01 | 4 |

The critical technology for wide field-of-view double Fourier interferometry is already mature. Detector arrays such as those aboard the Spitzer Space Telescope (IRAC and IRS instruments), the upcoming WISE MIDEX mission (http://www.astro.ucla.edu/~wright/WISE/), and their planned successors (http://safir.gsfc.nasa.gov/docs/ISMDWG_final.pdf) are well suited for this application, both in sensitivity and pixel count. The moving scan mechanism in the Composite Infrared Spectrometer (CIRS) on the Cassini mission provides ~10 cm scan range. FTS scan mechanism technology has extensive heritage in space (http://www.ngst.nasa.gov/public/unconfigured/doc_0517/rev_01/CSA_IFTS.pdf).

Caveat:

The Double Fourier method is best suited for low to moderate spectral resolution. Resolution in the R = $10^4$ –$10^6$ range is attainable with the double Fourier method, but a long delay line stroke is required, and the sensitivity is poorer than that available via dispersive methods and may be inadequate for certain sources and spectral lines.

# Appendices

# Appendix A
# List of Contributors

The editor is grateful to acknowledge the participation of the TPF Science Working Group, listed in Table 3, and contributions from selected participants of the workshop, listed in Table 4. This document was put together with the assistance of Peter Lawson (Jet Propulsion Laboratory).

**Table 3.** Science Working Group Subgroup

| Name | Email Address |
|---|---|
| Charles Beichman | chas@ipac.caltech.edu |
| William Danchi | wcd@iri1.gsfc.nasa.gov |
| Eric Gaidos | gaidos@hawaii.edu |
| Sara Heap | hrsheap@hrs.gsfc.nasa.gov |
| Tony Hull | tonyhull@jpl.nasa.gov |
| Kenneth J. Johnston | kjj@astro.usno.navy.mil |
| Steve Kilston | skilston@ball.com |
| Marc J. Kuchner | mkuchner@astro.princeton.edu |
| Doug Lin | lin@ucolick.org |
| René Liseau | rene@astro.su.se |
| Jonathan I. Lunine | jlunine@lpl.arizona.edu |
| Charley Noecker | mcnoecke@ball.com |
| Sara Seager | seager@dtm.ciw.edu |
| Eugene Serabyn | eserabyn@huey.jpl.nassa.gov |
| David Spergel | dns@astro.princeton.edu |
| William Sparks | sparks@stsci.edu |
| Karl Stapelfeldt | krs@exoplanet.jpl.nasa.gov |
| Huub Rottgering | rottgeri@strw.leidenuniv.nl |
| Ted von Hippel | ted@astro.as.utexas.edu |
| Bertrand Mennesson | bmenness@jpl.nasa.gov |

**Table 4.** Other Contributors

| Name | Email Address |
|---|---|
| Rachel Akeson | rla@ipac.caltech.edu |
| Ron Allen | rjallen@stsci.edu |
| Carol Ambruster | carol.ambruster@villanova.edu |
| Gary Bernstein | garyb@physics.upenn.edu |
| Chuck Bowers | bowers@stis.gsfc.nasa.gov |
| Thomas M, Brown | tbrown@stsci.edu |
| James Breckinridge | James.B.Breckinridge@jpl.nasa.gov |
| Robert Brown | rbrown@stsci.edu |
| Simon DeDeo | simon@astro.princeton.edu |
| Orsola De Marco | orsola@amnh.org |
| Olivier Dore | Olivier@astro.princeton.edu |
| Harry Ferguson | ferguson@stsci.edu |
| Andrea Ghez | ghez@astro.ucla.edu |
| Amir Give'on | agiveon@princeton.edu |
| Mustapha Ishak-Boushaki | mishak@astro.princeton.edu |
| Mario Juric | mjuric@astro.princeton.edu |
| Jeremy Kasdin | jkasdin@princeton.edu |
| Ben Lane | blane@mit.edu |
| David Leisawitz | leisawitz@stars.gsfc.nasa.gov |
| Douglas Lisman | p.d.lisman@jpl.nasa.gov |
| Milos Milosavljevic | milos@tapir.caltech.edu |
| Mike Rich | rmr@astro.ucla.edu |
| Doug Richstone | dor@umich.edu |
| Adam Riess | ariess@stsci.edu |
| Joop Schaye | schaye@ias.edu |
| Stuart Shaklan | stuart.shaklan@jpl.nasa.gov |
| Ed Sirko | esirko@astro.princeton.edu |
| Michael Strauss | strauss@astro.princeton.edu |
| Steve Unwin | stephen.unwin@jpl.nasa.gov |
| Bob Vanderbei | rvdb@princeton.edu |
| Dejan Vinkovic | dejan@ias.edu |
| Nevin Weinberg | nnw@tapir.caltech.edu |
| Bruce Woodgate | woodgate@stars.gsfc.nasa.gov |
| Harold Yorke | Harold.Yorke@jpl.nasa.gov |

# Appendix B
# Mission Parameters

Here are the parameters of the two missions assumed in this report.

## TPF-C Mission Parameters

*Virginia Ford, Tony Hull, Stuart Shaklan, and Karl Stapelfeldt*

PRIMARY MIRROR:

One unobscured 8.0m × 3.5m elliptical monolithic primary mirror.

WAVELENGTH COVERAGE:

Nominal value: 0.5 – 0.8 μm to achieve terrestrial planet detection and characterization goals
Trivial upgrade: 0.35 – 1.05 μm is accessible with silver coatings and CCD detector
Modest upgrade: none
Major upgrades: 1) access down to ~0.1 μm (TBR) conceivable if first 4–6 mirrors were Al coated, and separate camera were provided. Significant risk of degrading high-contrast performance must be evaluated since the stability, uniformity, and reflectance of Al is not optimal for primary science. Separate detector would likely be required for UV. 2) access out to 1.7 μm if second camera is provided with doped HgCdTe detector. Warm optics will probably prevent working longward of 1.7 μm.

ANGULAR RESOLUTION:

Nominal value: 13 mas ×µ 29 mas at 0.5 µm for minimum mission.
Value selected to allow access to 35 stars outside 4 lamda/D radius

Trivial upgrade: 13 mas at 0.5 µm in both axes by observing each target at multiple roll angles spread over 90 degrees to build up isotropic resolution

Modest upgrade: none

Major upgrade: none that can be contemplated within primary mission resources

POINTING ACCURACY:

Nominal value: 1 mas RMS accuracy is required to maintain star image on coronagraph occulting spot. Achieved with ~0.1" spacecraft body pointing and fine guiding to 1 mas (in tip/tilt only) internal to the coronagraph. Bright central target star with V < 7 provides guiding signal for a fine steering mirror. Images at a large distance off-axis in the coronagraphic camera could be smeared by uncontrolled spacecraft roll angle drift. Images in any second instrument will be smeared by 0.1" RMS boresight drift plus the roll angle drift.

Trivial upgrade: Full guiding performance should be possible with fainter guidestars, perhaps down to V = 16 (TBR), and guiding with degraded pointing performance on still fainter targets.

Modest upgrade: Add additional guiding camera to view a second star off axis, to control spacecraft roll drift that would degrade off-axis imaging at the edge of any extended FOV in the main coronagraphic camera. Still requires V < 16 (TBR) guidestar to be available adjacent to the desired target field.

Major upgrade: Install 2–3 Fine Guidance Sensors and precision reaction wheels, a la HST. Control spacecraft body pointing to 1 mas using signal from V < 16 guidestars, independent of the coronagraphic camera.

IMAGING FIELD OF VIEW:

Nominal value: The current baseline includes a detector with ~ 500 pixels covering a field 5 arcsec wide.

Trivial upgrade: Might be extended to 1–2 arcmin in primary science camera with an optical redesign

Modest upgrade: none

Major upgrade: Provide separate wide-field imaging camera, off axis, and covering > 10 square arcmin. Would drive large increase in number of focal planes, detector cooling requirements, on-board data storage, downlink data rate, spacecraft pointing requirements.

HIGH CONTRAST OUTER WORKING ANGLE:

Nominal value: OWA = 1.13" at 0.5 µm, assuming anamorphic mptics and deformable mirrors with 96 × 96 actuators. At angles > OWA, where the DM has no ability to control scattered light, the high quality of the optics enables detections 15.5 magnitudes fainter than the target star. This detection limit is expected to improve as roughly the square of the angular distance from the target star for larger angles, with an instrument limit 25 magnitudes fainter than the target star set by incoherent stray light in the system.
Trivial upgrade: none
Modest upgrade: none
Major upgrade: 1.51" × 1.51” at V band using a 128 × 128 deformable mirror

SPECTROSCOPIC CAPABILITY:

Nominal value: Spectral resolution of 70, minimum needed to reliably detect O2 A band at 0.77 µm. Integral field unit used to simultaneously sense the planet and the wavelength-dependent speckle pattern; spectral target must lie within 1.1" of the bright star used for guiding.
Trivial upgrade: none
Modest upgrade: Double the format of IFU backend detector to allow 4× improved spectral resolutuion, or 2× larger IFU field of view
Major upgrade: Provide separate spectrograph instrument with 200 > R > 10,000.

THROUGHPUT:

Nominal value: V = 30 point source target is detected to S/N = 4 in 7500 sec.

# TPF-I Mission Parameters

*Bertrand Mennesson, Gene Serabyn, and Huub Rottgering*

PRIMARY MIRRORS:

Four 3.5m circular mirrors.

WAVELENGTH COVERAGE:

Nominal value: 6 – 20 μm to achieve terrestrial planet detection and characterization goals
Trivial upgrade: 5 – 28 μm with a Si:As BIB detector
Modest upgrade: none
Major upgrades: 1) access down to ~3 μm

ANGULAR RESOLUTION:

Nominal value:
The interferometer has four free-flying telescopes in a chopped dual-Bracewell configuration with baselines 15 m – 1 km.

Closest detectable planet: ($\lambda$/2b) = 10 mas at 10 μm, baseline of 100m
Imaging resolution: ($\lambda$/b) = 20 mas at 10 μm, baseline of 100 m

Trivial upgrade: none
Modest upgrade: none
Major upgrade: With 3 μm capability, $\lambda$/2b can reach 0.3 mas

IMAGING FIELD OF VIEW:

Nominal value: R$\lambda$/b, where R is the spectral resolution. The nulled light goes through fibers, so the FOV is a single mode.
Trivial upgrade: Increase R.
Modest upgrade: none
Major upgrades: Fizeau Beam Combiner, Double Fourier Beam Combiner (see below)

HIGH CONTRAST OUTER WORKING ANGLE: same as field of view

SPECTROSCOPIC CAPABILITY:

Nominal value: Spectral resolution R = 20 to 75
Trivial upgrade: none
Modest upgrade: Add a grating for R ~ 1000, bright objects only

Major upgrade: Fourier Transform Spectrometer for R = 100,000, bright objects only

THROUGHPUT:

Nominal value: Sensitivity ≈ .4 μJy/pix at 12 μm in $10^5$ s
Detectable Surface Brightness scales with b
Integration time is a complex function of source brightness, complexity, an desired (u,v) coverage. Overall time spent on imaging determined by outcome of planet detection and characterization phase. Sensitivity assumes a phase reference with K<17, assuming ~10s of coherence time in space.

SKY COVERAGE:

Nominal value: Fringe tracker requires targets with K<17
Trivial upgrade: none
Modest upgrade: none
Major upgrade: Dual field interferometry over 1’ would allow 100% sky coverage

OBSERVING MODES:

Nominal: Nulling
Trivial: $V^2$
Modest upgrades: phase closures
Major upgrade: absolute phase referencing

# Appendix C
# Acronyms

| | |
|---|---|
| ACS | Advanced Camera for Surveys |
| AGB | Asymptotic Giant Branch |
| AGN | Active Galactic Nuclei |
| ALMA | Atacama Large Millimeter Array |
| AU | Astronomical Unit |
| CCD | Charge-Coupled Device |
| CIRS | Composite Infrared Spectrometer (for Cassini) |
| CMOS | Complementary Metal Oxide Silicon |
| ESSENCE | Equation of State: SupErNovae trace Cosmic Expansion |
| FIRES | Faint InfraRed Extragalactic Survey |
| FOV | Field of View |
| FTS | Fourier Transform Spectrometer |
| FWHM | Full-Width at Half Maximum |
| GALEX | Galaxy Evolution Explorer |
| GOODS | Great Observatories Origins Deep Survey |
| GSFC | Goddard Space Flight Center |
| HST | Hubble Space Telescope |
| IFS | Integral Field Spectograph |
| IFU | Integral Field Unit |
| IGM | Inter-Galactic Medium |
| IRAC | Infrared Camera Array |
| ISM | Inter-Stellar Medium |
| JWST | James Webb Space Telescope |
| LSST | Large Synoptic Survey Telescope |
| MIDEX | Medium-class Explorers |
| NASA | National Aeronautics and Space Administration |
| PSF | Point Spread Function |
| QSO | Quasi-Stellar Object |
| RMS | Root Mean Square |
| SNAP | SuperNova/Acceleration Probe |
| STIS | Space Telescope Imaging Spectrograph |
| TBR | |
| TPF | Terrestrial Planet Finder |
| TPF-C | Terrestrial Planet Finder Coronagraph |
| TPF-I | Terrestrial Planet Finder Interferometer |
| UC | University of California |
| UCO | University of California Observatories |
| URL | Universal Resource Locator |
| USNO | United States Naval Observatory |
| UV | Ultraviolet |
| WIIT | Wide-field Imaging Interferometer Testbed |
| WISE | Wide-field Infrared Survey Explorer |
| ZPD | Zero Path Difference |

# Appendix D
# Figure Notes and Copyright Permissions

Figure 1 – Original by S. Dedeo (Princeton University) and E. Sirko (Princeton University) for this publication.

Figure 2 – Original by Ben Lane (MIT) and Sally Heap (Goddard Space Flight Center) for this publication.

Figure 3 – Reprinted, by permission of the publisher, from *Astrophysical Journal* **158**, 809 (1969)

Figure 4 – Reprinted, by permission of the publisher, from *Astrophysical Journal* **513**, 879 (1999)

Figure 5 – Reprinted, by permission of the publisher, from *Astrophysical Journal* **618** (2005)

Figure 6 – Original by Marc Kuchner (Princeton University) for this publication.

Figure 7 – Original by Dejan Vinkovic (Institute for Advanced Study) for this publication.

Figure 8 – Original by Nevin Weinberg (Caltech) and Andrea Ghez (UCLA) for this publication.

Figure 9 – Original by Tom Brown (Space Telescope Science Institute) for this publication.

Figure 10 – Reprinted, by permission of the publisher, from *Astrophys. J.* **476**, 327 (1997)

Figure 11 – Original by Huub Rottgering (Leiden University) for this publication.

Figure 12 – Original by Huub Rottgering (Leiden University) for this publication.

Figure 13 – Reprinted, by permission of the publisher, from *Astrophys. J.* **454**, L175 (1995)

Figure 14 – The Hubble Ultra Deep Field (NASA).

Figure 15 – Original by Mustapha Ishak-Boushaki for this publication.

Figure 16 – Original by Douglas Lisman for this publication.

Figure 17 – Original by Chuck Bowers and Bruce Woodgate for this publication.

Figure 18 – Original by Chuck Bowers and Bruce Woodgate for this publication.

Figure 19 – Original by Chuck Bowers and Bruce Woodgate for this publication.

Figure 20 – Original by David Leisawitz for this publication.

# Appendix E
# Further Reading

*Terrestrial Planet Finder — News*
Edited by R. Jackson (Jet Propulsion Laboratory)
http://planetquest.jpl.nasa.gov/TPF/tpf_news.cfm

NASA Research Opportunities
http://research.hq.nasa.gov/research.cfm

*Astronomy and Astrophysics in the New Millennium*
National Academies Press (2001)
http://www.nas.edu/bpa2/nsindex.html

*Precursor Science for the Terrestrial Planet Finder*,
P.R. Lawson, S.C. Unwin, and C.A. Beichman
Jet Propulsion Laboratory, Pasadena, CA: JPL Pub 04-014 (2004)
http://planetquest.jpl.nasa.gov/documents/RdMp273.pdf

*TPF- A NASA Origins Program to Search for Habitable Planets*
Edited by C.A. Beichman, N.J. Woolf, and C.A. Lindensmith
Jet Propulsion Laboratory, Pasadena, CA: JPL Pub 99-3 5/99 (1999)
http://planetquest.jpl.nasa.gov/TPF/tpf_book/index.cfm